\documentclass[sigconf]{acmart}

\usepackage{booktabs} 
\usepackage{multirow}
\usepackage{latexsym}
\usepackage{graphicx} 
\usepackage{algorithm}
\usepackage{algorithmic}
\usepackage{epsfig}
\usepackage{color}
\usepackage{amsmath}
\usepackage{graphicx,subfigure}
\usepackage{makecell}
\usepackage{textcomp}
\usepackage{url}
\usepackage{flushend}

\copyrightyear{2021} 
\acmYear{2021} 
\setcopyright{acmcopyright}
\acmConference[SIGIR '21] {Proceedings of the 44th International ACM SIGIR Conference on Research and Development in Information Retrieval}{July 11--15, 2021}{Virtual Event, Canada.}
\acmBooktitle{Proceedings of the 44th International ACM SIGIR Conference on Research and Development in Information Retrieval (SIGIR '21), July 11--15, 2021, Virtual Event, Canada}
\acmPrice{15.00}
\acmISBN{978-1-4503-8037-9/21/07} 
\acmDOI{10.1145/3404835.3462861}

\author{Tao Qi$^1$, Fangzhao Wu$^2$, Chuhan Wu$^1$, Yongfeng Huang$^1$}

\affiliation{%
  \institution{$^1$Department of Electronic Engineering \& BNRist, Tsinghua University, Beijing 100084, China \\ $^2$Microsoft Research Asia, Beijing 100080, China}
} 
\email{{taoqi.qt,wufangzhao,wuchuhan15}@gmail.com,yfhuang@tsinghua.edu.cn}

\begin{document}
\title{Personalized News Recommendation with Knowledge-aware\\ Interactive Matching}


\begin{abstract}

The most important task in personalized news recommendation is accurate matching between candidate news and user interest.
Most of existing news recommendation methods model candidate news from its textual content and user interest from their clicked news in an independent way.
However, a news article may cover multiple aspects and entities, and a user usually has different kinds of interest.
Independent modeling of candidate news and user interest may lead to inferior matching between news and users.
In this paper, we propose a knowledge-aware interactive matching method for news recommendation.
Our method interactively models candidate news and user interest to facilitate their accurate matching.
We design a knowledge-aware news co-encoder to interactively learn representations for both clicked news and candidate news by capturing their relatedness in both semantic and entities with the help of knowledge graphs.
We also design a user-news co-encoder to learn candidate news-aware user interest representation and user-aware candidate news representation for better interest matching.
Experiments on two real-world datasets validate that our method can effectively improve the performance of news recommendation.

\end{abstract}



\maketitle

\section{INTRODUCTION}

Online news platforms such as Microsoft News, Apple News and News Break, have attracted a huge number of users to consume news information~\cite{okura2017embedding,wu2019ijcai}.
However, since massive new published news articles are collected by these platforms every day, users often have difficulties in finding the news information they need~\cite{wu2019npa,danzhu2019}.
Personalized news recommendation techniques, which aim to help users find their interested news, usually play an essential role in online news platforms to alleviate the information overload of users~\cite{wang2018dkn,an2019neural}.
Thus, the study on personalized news recommendation has attracted much attention from both academia and industry~\cite{bansal2015content,an2019neural,kompan2010content,khattar2018weave,zheng2018drn,wuuser,liu2010personalized,wu2020ptum}.

Accurate matching between user interest and candidate news is critical for personalized news recommendation~\cite{wang2020fine,wang2018dkn}.
Existing methods usually model candidate news from its textual information and infer user interest from user's click history in an independent way~\cite{okura2017embedding,wu2019neuralc}.
For example, \citet{wu2019npa} learned news representations via a word-level personalized attention network and learned user interest representations via a user-level personalized attention network, independently.
They further performed interest matching via the inner product of user interest representation and candidate news representation.
However, a candidate news article may contain multiple aspects and entities~\cite{wu2019ijcai,liu2020kred}, and a user may have multiple interests~\cite{wang2018dkn}.
Thus, independent modeling of candidate news and user interest may be inferior for the interest matching~\cite{wang2020fine}.

\begin{figure}
    \centering
    \resizebox{0.5\textwidth}{!}{
    \includegraphics{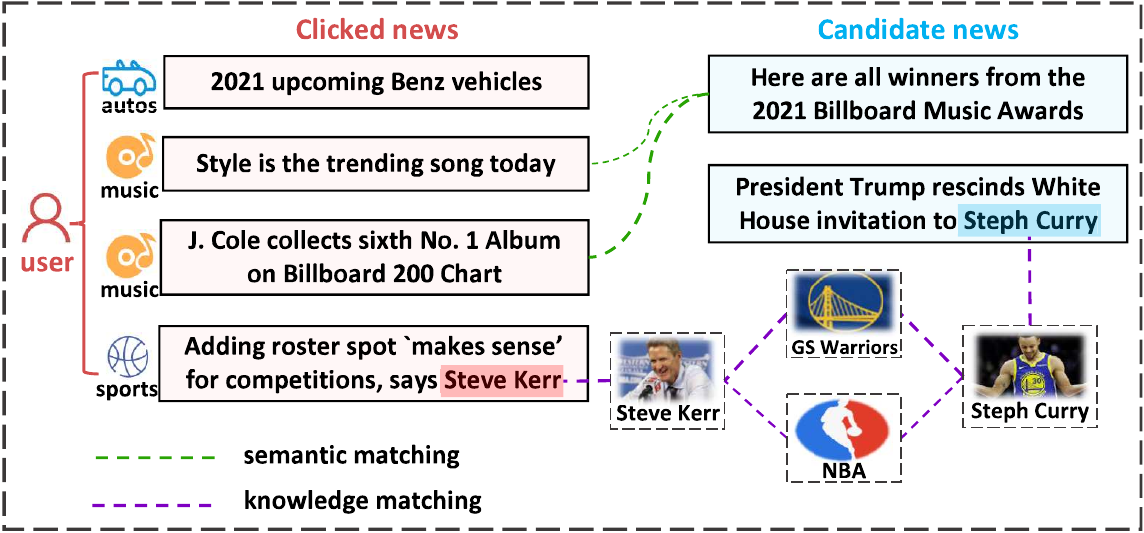}
    }
    \caption{An example user with four clicked news, and two example candidate news for recommendation.}
    \label{fig.motivation}
\end{figure}

In this paper, we explore to better model the relatedness between candidate news and user interests for accurate interest matching.
Our paper is motivated by the following observations.
First, a candidate news may cover different aspects and entities, and a user may have multiple interests.
For example, the 2nd candidate news in Fig.~\ref{fig.motivation} is related to a basketball star and a politician, and covers several entities, e.g., ``Stephen Curry'' and ``Donald Trump''.
Besides, the example user in Fig.~\ref{fig.motivation} is interested in multiple areas such as autos, music and sports.
The 2nd candidate news can only match a specific user interest, i.e., sports, and the user may be only interested in a single entity in the 2nd candidate news, i.e., ``Stephen Curry''.
Thus, it is inferior for matching user interest with candidate news if they are independently modeled.
Second, semantic matching of candidate news and clicked news can help perform interest matching more accurately.
For instance, the 2nd clicked news also has semantic relatedness with the 1st candidate news since both of them are related to music.
The 3rd clicked news has semantic relatedness with the 1st candidate news since they mention the same event.
Based on these semantic relatedness, we can infer the user may be interested in the 1st candidate news.
Third, with the help of knowledge graphs, the knowledge matching between entities in clicked news and candidate news is also informative for understanding user interest in candidate news.
For example, the entity ``Steve Kerr'' in the 4th clicked news has inherent relatedness with the entity ``Stephen Curry'' in the 2nd candidate news since the former and the latter is the player and coach of the ``Warriors'' team of the ``NBA'' competition, respectively.
According to the knowledge matching, we can infer the user may have interest in the 2nd candidate news.
Thus, exploiting the relatedness between clicked news and candidate news in both semantic and knowledge levels is beneficial for interest matching.

In this paper, we propose a knowledge-aware interactive matching framework for personalized news recommendation (named \textit{KIM}).
Our method can interactively model candidate news and user interest to learn candidate news-aware user interest representation and user-aware candidate news representation to match user interest and candidate news more accurately.
In the framework, we propose a knowledge co-encoder to model user interest in candidate news from the relatedness between entities in clicked news and candidate news with the help of knowledge graphs.
More specifically, we first propose a graph co-attention network to learn representations of entities from the knowledge graph by selecting and aggregating their neighbors that are informative for interest matching.
We further propose to use an entity co-attention network to interactively learn knowledge-based representations of both clicked news and candidate news by capturing relatedness between their entities.
Moreover, we also propose a semantic co-encoder to interactively learn semantic-based representations for user's clicked news and candidate news by modeling semantic relatedness between their texts.
The unified representation of news is formulated as the aggregation of its knowledge- and semantic-based representation.
In addition, we further propose a user-news co-encoder to build candidate news-aware user interest representation and user-aware candidate news representation from representations of clicked news and candidate news to better model user interest in candidate news.
Finally, the candidate news is ranked based on the relevance between representations of candidate news and user interest.
We conduct extensive experiments on two real-world datasets and show that our method can effectively improve the performance of news recommendation and outperform other baseline methods.

\section{RELATED WORK}


Personalized news recommendation is an important task for online news services~\cite{das2007google,lin2014personalized} and has been widely studied in recent years~\cite{konstan1997grouplens,wang2011collaborative,wutanr,lian2018towards,wu2019neurald,qi2020privacy,wu2020sentirec,wu2020fairness}. 
Existing methods usually model candidate news from its content and model user interest from clicked news independently, and then match candidate news and user interests based on their relevance~\cite{wu2019neuralc,wu2019ijcai,wang2018dkn,wu2020ccf,ge2020graph}.
For example, \citet{okura2017embedding} represented candidate news from its bodies via an auto-encoder and represented user interest from user's click history via a GRU network, independently.
They further matched user interest and candidate news based on the dot product between their representations.
\citet{wu2019neuralc} adopted a multi-head self-attention network to model candidate news from its title and another multi-head self-attention network to model user interest from user's click history.
\citet{liu2020kred} proposed to learn knowledge-based candidate news representation from entities in news title and their neighbors on knowledge graphs and learn user interest representation from user's clicked news via an attention network.
Besides, these methods also performed interest matching via the inner product of user interest representation and candidate news representation.
In general, a candidate news may cover multiple aspects and entities~\cite{wu2019ijcai,liu2020kred}, and a user may have multiple interests~\cite{wang2018dkn}.
Only a part of candidate new aspects and user interests are useful for matching user interest with candidate news.
However, these methods model candidate news and user interest independently, which may be inferior for the further interest matching.
Different from these methods, in the \textit{KIM} method we propose a knowledge-aware interactive matching framework to interactively model candidate news and user interest with the consideration of their relatedness, which can better match user interest with candidate news.

Some methods model user interest in a candidate-aware way~\cite{wang2018dkn,danzhu2019}.
For example, \citet{wang2018dkn} obtained news representations from aligned words and entities in news titles via a multi-channel CNN network.
Besides, they applied a candidate-aware attention network to learn user interest representation by aggregating clicked news based on their relevance with candidate news.
They further used a dense network to model the relevance of user interest and candidate news from their representations.
\citet{danzhu2019} proposed to learn news representations from words and entities in news titles via multiple CNN networks and learn user interest representations from historical clicks via a LSTM network and an candidate-aware attention network.
They matched user interest and candidate news based on the cosine similarity of their representations.
In fact, candidate news may contain multiple aspects and entities~\cite{wu2019ijcai,liu2020kred} and only a part of them may match user interest.
However, these methods model candidate news without the consideration of the target user, which maybe inferior for further matching user interest with candidate news.
Different from these methods, our \textit{KIM} method models candidate news with the consideration of target user.
In addition, these methods model clicked news and candidate news without the consideration of their relatedness, which may also be suboptimal for further measuring relevance between candidate news and user interest inferred from clicked news.
Different from these methods, \textit{KIM} can interactively learn representations of both clicked news and candidate news for better interest matching.

\begin{figure}
    \centering
    \resizebox{0.44\textwidth}{!}{
    \includegraphics{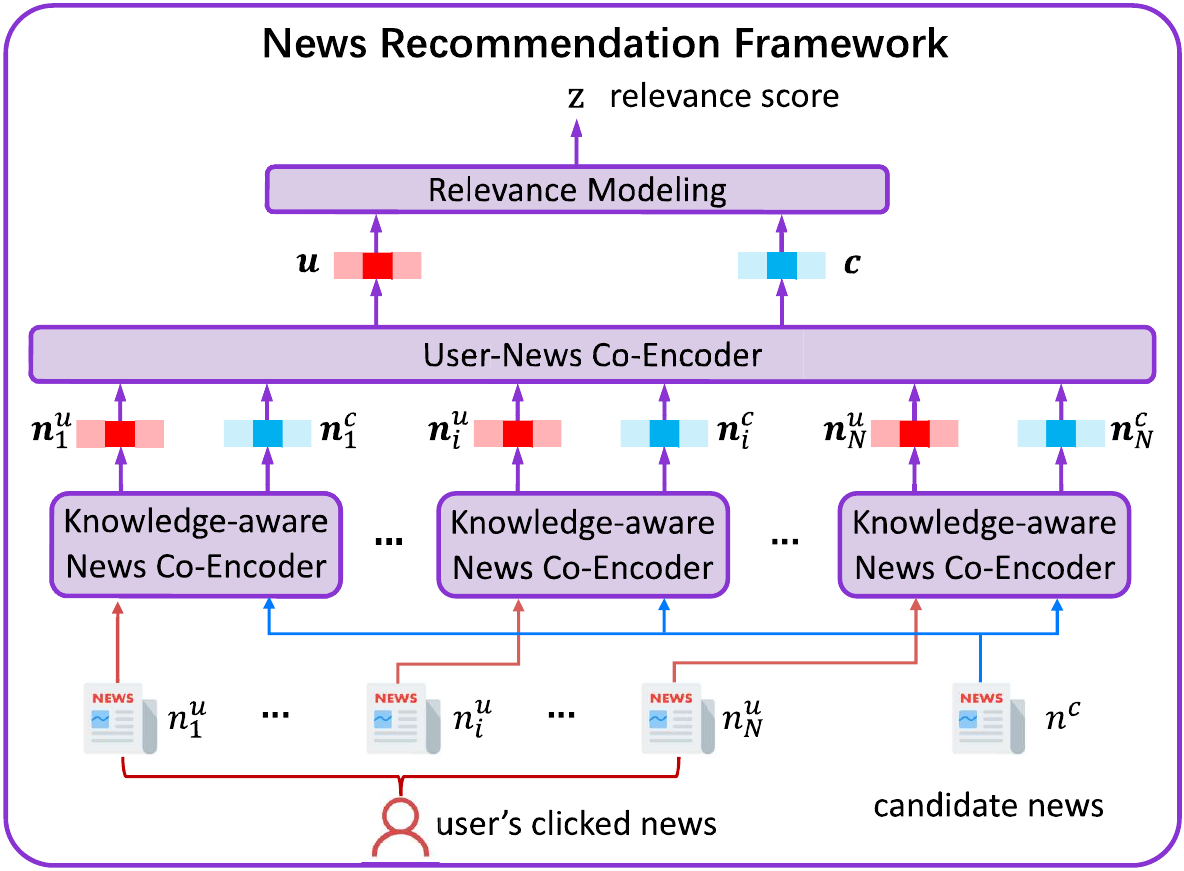}
    }
    \caption{The overall framework of \textit{KIM}.}
    \label{fig.framework}
\end{figure}

\section{Methodology}

We first introduce the problem definition of personalized news recommendation.
Next, we introduce our knowledge-aware interactive matching framework for personalized news recommendation (named \textit{KIM}) in detail.




\subsection{Problem Formulation}
Given a user $u$ and a candidate news $n^c$, we need to compute the relevance score $z$ measuring the interest of user $u$ in the content of candidate news $n^c$.
Then different candidate news are ranked and recommended to user $u$ based on their relevance scores.
The user $u$ is associated with the set of his/her clicked news.
Each news $n$ is associated with its texts $T$ and entities $E$ in its texts.
Besides, there is a knowledge graph $\mathcal{G}$ used to provide the relatedness between entities.
It contains entities and relations between entities.
Each entity $e$ in $\mathcal{G}$ is associated with its embeddings $\textbf{e}$ pre-trained based on the knowledge graph.
In our method, we only utilize the links between entities to represent their relatedness and do not utilize the specific relations (e.g., located\_at). 


\subsection{Framework of KIM}

In this section, we introduce the news recommendation framework of \textit{KIM}, which can interactively model candidate news and user interest for better interest matching.
As illustrated in Fig.~\ref{fig.framework}, \textit{KIM} contains two major modules.
The first one is a \textit{knowledge-aware news co-encoder}, which interactively learns the knowledge-aware representations of a user's clicked news and the candidate news by capturing their relatedness at both semantic and knowledge levels.
The second one is a \textit{user-news co-encoder}, which interactively learns candidate news-aware user interest representation $\textbf{u}$ and user-aware candidate news representation $\textbf{c}$ from the representations of user's clicked news and candidate news generated by the \textit{knowledge-aware news co-encoder}.
Finally, we match candidate news with user interest based on the relevance between the candidate news-aware user interest representation and user-aware candidate news representation.
Next, we introduce each module in detail.

\begin{figure}
    \centering
    \resizebox{0.46\textwidth}{!}{
    \includegraphics{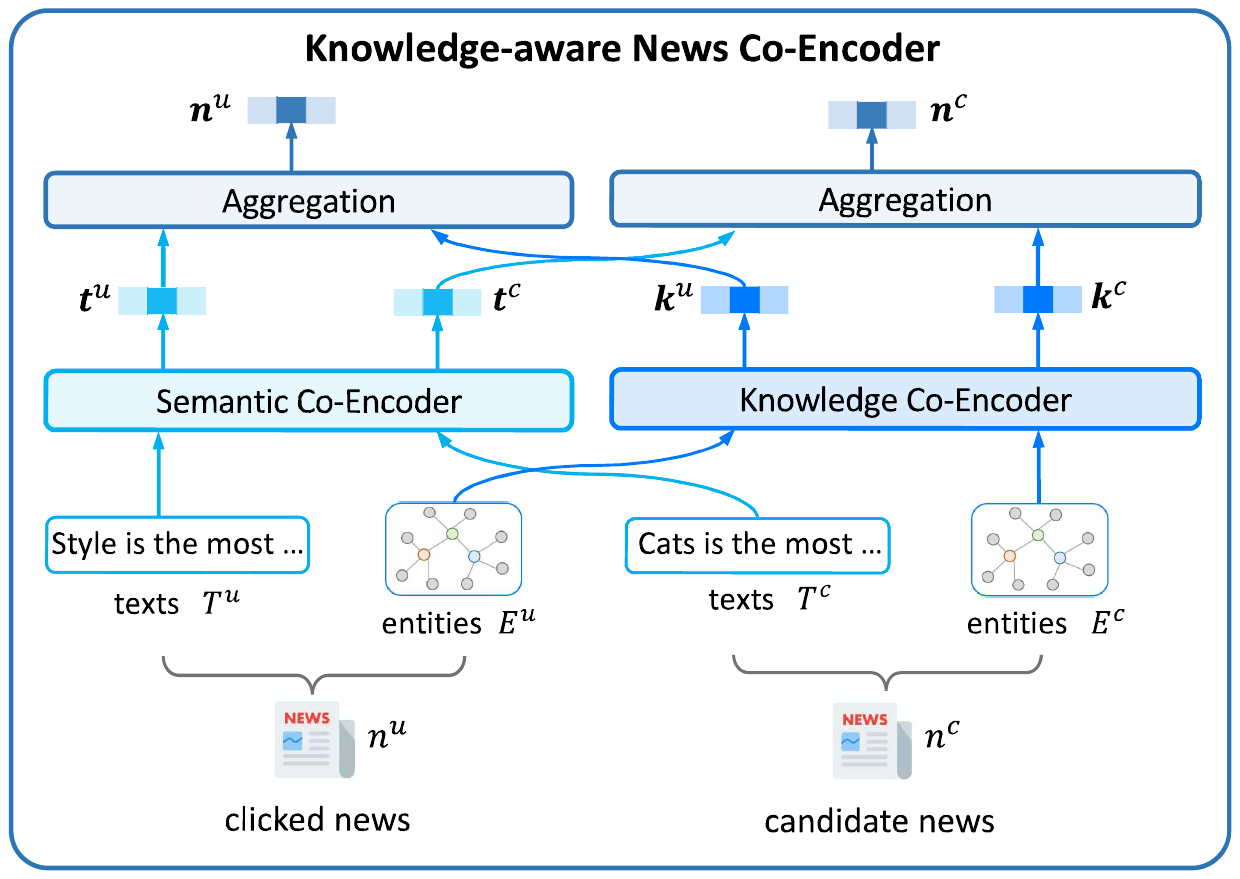}
    }
    \caption{The knowledge-aware news co-encoder in \textit{KIM}.}
    \label{fig.news_encoder}
\end{figure}

\subsection{Knowledge-aware News Co-Encoder}

\begin{figure*}
    \centering
    \resizebox{0.95\textwidth}{!}{
    \includegraphics{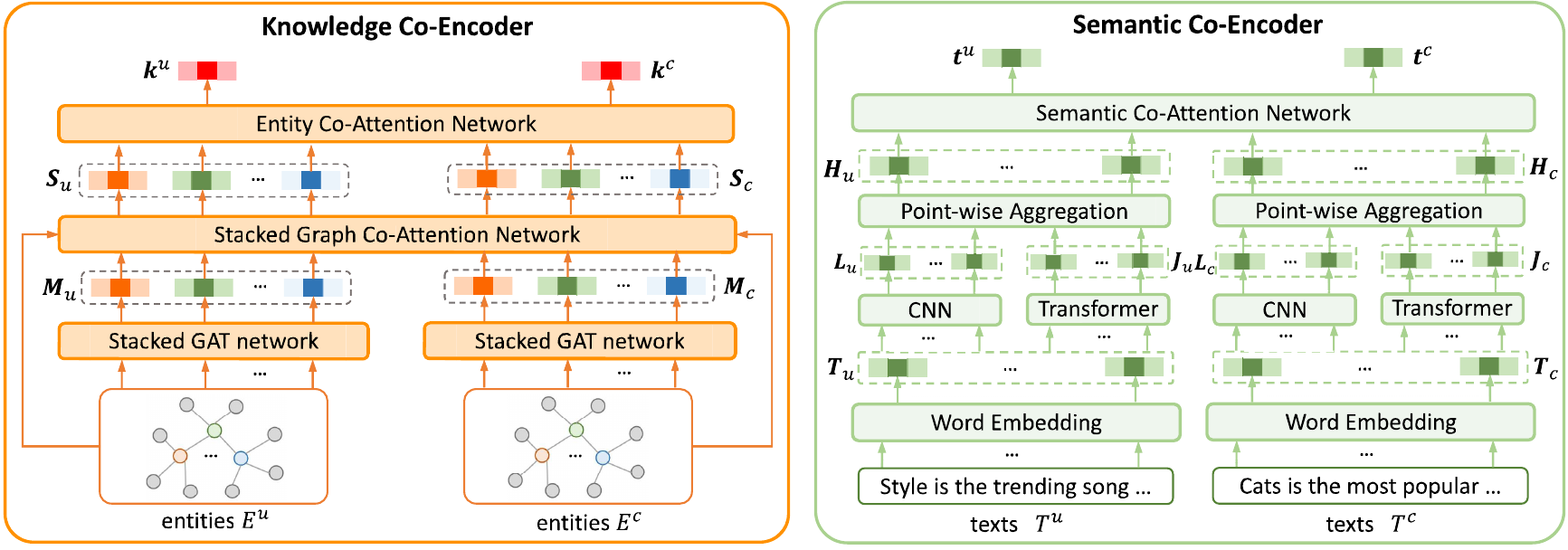}
    }
    \caption{The architecture of the knowledge co-encoder and semantic co-encoder.}
    \label{fig.models}
\end{figure*}

\begin{figure}
    \centering
    \resizebox{0.48\textwidth}{!}{
    \includegraphics{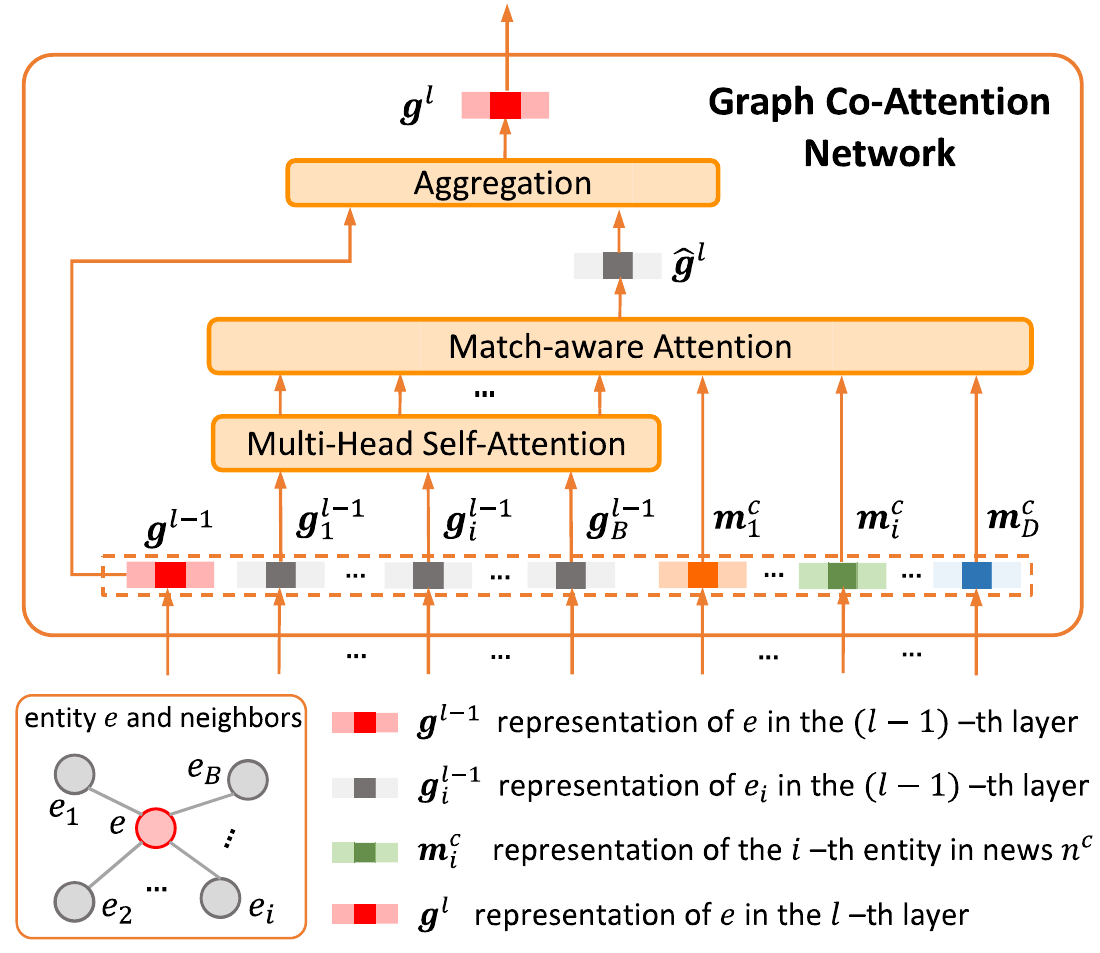}
    }
    \caption{Graph co-attention network (GCAT).}
    \label{fig.gcat}
\end{figure}

In this section, we introduce the framework of the \textit{knowledge-aware news co-encoder}, which interactively learns representations of a user's clicked news $n^u$ and candidate news $n^c$ from their texts and entities in texts.
As shown in Fig.~\ref{fig.news_encoder}, it contains three sub-modules.
The first one is a \textit{knowledge co-encoder} (denoted as $\Phi_k$), which interactively learns knowledge-based representations $\textbf{k}^u\in \mathcal{R}^{ d_k}$ and $\textbf{k}^c\in \mathcal{R}^{ d_k}$ for clicked news $n^u$ and candidate news $n^c$ from the relatedness between their entities based on the knowledge graph:
\begin{equation}
    [\textbf{k}^u, \textbf{k}^c] = \Phi_k(E^u,E^c),
\end{equation}
where $d_k$ denotes knowledge-based news representation dimensions, $E^u$ and $E^c$ denote entities in news $n^u$ and $n^c$ respectively.
The second one is a \textit{semantic co-encoder} (denoted as $\Phi_t$), which interactively learns semantic-based representations $\textbf{t}^u \in\mathcal{R}^{d_t} $ and $\textbf{t}^c \in\mathcal{R}^{d_t}$ for news $n^u$ and $n^c$ to model user interests in candidate news from the semantic relatedness between their texts:
\begin{equation}
    [\textbf{t}^u,\textbf{t}^c] = \Phi_t(T^u,T^c),
\end{equation}
where $d_t$ denotes semantic-based news representation dimensions, $T^u$ and $T^c$ denote texts of news $n^u$ and $n^c$ respectively.
Finally, we project the knowledge- and semantic-based representation of the same news to learn the unified news representation:
\begin{equation}
\textbf{n}^u = \textbf{P}_n [\textbf{t}^u; \textbf{k}^u], \quad \quad \textbf{n}^c = \textbf{P}_n [\textbf{t}^c; \textbf{k}^c],
\end{equation}
where $\textbf{n}^u  \in\mathcal{R}^{d_n} $ denotes the knowledge-aware representation of user's clicked news $n^u$, $\textbf{n}^c  \in\mathcal{R}^{d_n}$ denotes the corresponding knowledge-aware representation of candidate news $n^c$, $d_n$ denotes news representation dimensions, $[\cdot;\cdot]$ denotes the concatenation operation, and $\textbf{P}_n  \in\mathcal{R}^{d_n \times (d_t+d_k) }$ is the trainable projection matrix.

\subsubsection{Knowledge Co-Encoder}
We introduce the proposed \textit{knowledge co-encoder}, which interactively learns the knowledge-based representations of user's clicked news $n^u$ and candidate news $n^c$.
It aims to better represent these news for interest matching from relatedness between entities $E^u$ and $E^n$ in user's clicked news and candidate news with the help of the knowledge graph $\mathcal{G}$.
As shown in Fig.~\ref{fig.models}, it contains three components.
First, to summarize the information for each entity in $E^u$ or $E^c$ from their neighbors within $K$ hops, we first utilize a graph attention (GAT) network~\cite{velivckovic2017graph} stacked $K$ layers to learn their representations, which are denoted as $\textbf{M}_u = \{\textbf{m}^u_i\}_{i=1}^D\in \mathcal{R}^{d_k\times D} $ and $\textbf{M}_c = \{\textbf{m}^c_i\}_{i=1}^D \in \mathcal{R}^{d_k\times D} $ respectively, where $D$ is the number of entities in news.


The second one is a stacked graph co-attention (GCAT) network proposed in this paper.
In general, an entity usually has rich relatedness with multiple entities on the knowledge graph~\cite{wan2008single,hamaguchi2017knowledge}.
Besides, relatedness among entities usually provides different informativeness to model the relatedness between clicked news and candidate news for interest matching.
For example, given a clicked news ``Style is the trending song today.'' and a candidate news ``The movie Cats is the most popular movie in the Netflix.'', the entity ``Movie Cats'' has many neighbor entities on knowledge graphs, such as its director ``James'', chief actor ``Hooper'', chief actress ``Taylor'' and so on.
Only the entity ``Taylor'' is informative for modeling the relatedness between clicked news and candidate news since it is also the singer of the entity ``Song Style'' in clicked news.
To better select informative relatedness between entities for matching candidate news with user interest, we propose a graph co-attention network (GCAT) stacked $K$ layers to learn match-aware representations for entities in news $n^u$ and $n^c$.
Take an entity $e$ in news $n^u$ as example, the $l$-th graph co-attention network shown in Fig.~\ref{fig.gcat} learns its representation by aggregating representations of its neighbors guided by entities in news $n^c$.
More specifically, we first apply a multi-head self-attention network~\cite{vaswani2017attention} to the representations of its neighbor entities generated by the $(l-1)$-th GCAT network\footnote{The input of the $1$-th GCAT network are the initialized embeddings of each entity.} to model the conceptual relatedness between different neighbor entities.
Next, we propose a match-aware attention network to aggregate neighbor entities of entity $e$ based on their relevance with entities in news $n^c$ measured by a relevance matrix $\textbf{I}_u \in \mathcal{R}^{D\times B}$:
\begin{equation}
    \textbf{I}_u = \textbf{M}_c^T \textbf{W}^c_c \hat{\textbf{G}}_{l},
\end{equation}
where $\hat{\textbf{G}}_{l} =\{\hat{\textbf{g}}^l_i \}_{i=1}^B \in \mathcal{R}^{d_k\times B} $ denotes representations of neighbor entities generated by the self-attention network, $B$ denotes the number of neighbors, and $\textbf{W}^c_c\in \mathcal{R}^{d_k \times d_k} $ is trainable weights.
Then the attention vector $\textbf{v}^u \in \mathcal{R}^B $ of neighbor entities is calculated as:
\begin{equation}
    \textbf{v}^u = \textbf{q}_e^T \cdot \tanh(\textbf{W}^c_s\hat{\textbf{G}}^{l} +  \textbf{W}^c_h \textbf{M}_c f(\textbf{I}_u) ),
\end{equation}
where $f$ denotes the softmax activation which normalizes each column vector of the input matrix, $\textbf{q}_e \in \mathcal{R}^{d_q} $ denotes the trainable attention query, $d_q$ denotes its dimensions, $\textbf{W}^c_s \in \mathcal{R}^{d_q\times d_k} $ and $\textbf{W}^c_h \in \mathcal{R}^{d_q\times d_k}$ are trainable weights.
Then we aggregates neighbors of entity $e$ into a unified representation $\hat{\textbf{g}}^{l}\in \mathcal{R}^{d_k} $ :
\begin{equation}
    \hat{\textbf{g}}^{l} = \sum_{i=1}^B \lambda^u_i \hat{\textbf{g}}^{l}_i, \quad\quad \lambda^u_i = \frac{\exp(v^u_i)}{\sum_{j=1}^B \exp(v^u_j) }
\end{equation}
where $v^u_i$ is the $i$-th element of vector $\textbf{v}^u$ and $\lambda_i^u$ denotes the attention weight of the $i$-th neighbor entity.
Finally the representation $\textbf{g}^l\in \mathcal{R}^{d_k} $ of the entity $e$ generated by the $l$-th GCAT network is formulated as: $\textbf{g}^l = \textbf{P}_e  [\hat{\textbf{g}}^l; \textbf{g}^{l-1}],$
where $\textbf{P}_e \in \mathcal{R}^{d_k\times2d_k} $ is the projection matrix.
In this way, the GCAT network stacked $K$ layers can learn match-aware representations $\textbf{S}_u=\{\textbf{s}^u_i\}_{i=1}^D\in \mathcal{R}^{d_k\times D}$ for entities in user's clicked news by capturing the relatedness between their neighbors within $K$ hops and entities in candidate news, where $\textbf{s}^u_i$ is the representation of the $i-$th entity in clicked news $n^u$.
In a symmetrical way, we can learn the match-aware representations $\textbf{S}_c=\{\textbf{s}^c_i\}_{i=1}^D \in \mathcal{R}^{d_k\times D}$ of entities in candidate news from relatedness between their neighbors and entities in clicked news, where $\textbf{s}^c_i$ is the representation of the $i-$th entity in candidate news $n^c$.

The third one is an entity co-attention network.
Entities in clicked news and candidate news usually have different informativeness for interest matching.
For example, given a clicked news ``Style is the trending song in iTunes this week.'' and a candidate news ``The movie Cats is the most popular movie in the Netflix.'', the entity ``Song Style'' is more informative than the entity ``iTunes'' for matching user interest with candidate news since the entity ``Song Style'' has inherent relatedness with the entity ``Movie Cats'' in candidate news.
Thus, we apply an entity co-attention network to interactively learn knowledge-based representations for news $n^u$ and $n^c$ by capturing relatedness between their entities.
In detail, we first calculate an affinity matrix $\textbf{C}_e\in \mathcal{R}^{D\times D} $ to measure the relevance among entities in news $n^u$ and $n^c$:
\begin{equation}
    \textbf{C}_e = {\textbf{S}}_c^T \textbf{W}^k_c  {\textbf{S}}_u,
\end{equation}
where $\textbf{W}^k_c\in \mathcal{R}^{d_k\times d_k}$ is the trainable weights.
Then we calculate attention vectors $\textbf{a}^u,\textbf{a}^c \in \mathcal{R}^{D} $ of entities in news $n^u$ and $n^c$:
\begin{equation}
    \textbf{a}^{u} = \textbf{q}_k^T\cdot \tanh(\textbf{W}^k_s {\textbf{S}}_u + \textbf{W}^k_h {\textbf{S}}_c f(\textbf{C}_e)) ,
\end{equation}
\begin{equation}
    \textbf{a}^{c} = \textbf{q}_k^T\cdot \tanh(\textbf{W}^k_s {\textbf{S}}_c + \textbf{W}^k_h {\textbf{S}}_u  f(\textbf{C}^T_e)),
\end{equation}
where $\textbf{q}_k\in \mathcal{R}^{d_q}$ is the trainable attention query, and $\textbf{W}^k_s \in \mathcal{R}^{d_q\times d_k}$, $\textbf{W}^k_h \in \mathcal{R}^{d_q\times d_k}$ are trainable weights.
Finally we obtain knowledge-based representations $\textbf{k}^u \in\mathcal{R}^{d_k} $ and $\textbf{k}^c \in\mathcal{R}^{d_k} $ of clicked news and candidate news by aggregating their entities respectively:
\begin{equation}
    \textbf{k}^u = \sum_{i=1}^D \alpha^u_i {\textbf{s}}^u_i, \quad \quad \alpha^u_i = \frac{\exp(a^u_i)}{\sum_{j=1}^D \exp(a^u_j)},
\end{equation}
\begin{equation}
    \textbf{k}^c = \sum_{i=1}^D \alpha^c_i {\textbf{s}}^c_i, \quad \quad \alpha^c_i = \frac{\exp(a^c_i)}{\sum_{j=1}^D \exp(a^c_j)},
\end{equation}
where $\alpha^u_i$ and $\alpha^c_i$ denote the attention weight of the $i$-th entity in news $n^u$ and $n^c$ respectively.


\subsubsection{Semantic Co-Encoder}
As shown in Fig.~\ref{fig.models}, \textit{semantic co-encoder} interactively learns the semantic-based representations for user's clicked news $n^u$ and candidate news $n^c$.
It aims to better model user interests in candidate news from semantic relatedness between their texts ($T^u$ and $T^c$).
We first independently learn contextual representations for words in texts $T^u$ and $T^c$. 
More specifically, take texts $T^u$ as an example, we first convert it into an embedding vector sequence $\textbf{T}_u \in\mathcal{R}^{d_g\times M } $ via a word embedding layer, where $d_g$ denotes word embedding dimensions, and $M$ denotes the number of words in $T^u$.
Next, since both local and global contexts are important for semantic modeling~\cite{wu2019ijcai,wu2019neuralc}, we apply a CNN network~\cite{kim2014convolutional} and a transformer network~\cite{vaswani2017attention} to $\textbf{T}_u$ to learn both local- and global-contextual word representations respectively, i.e., $\textbf{L}_u \in\mathcal{R}^{d_t\times M}$ and  $\textbf{J}_u\in\mathcal{R}^{d_t\times M}$.
Then, we add the local- and global-contextual representations of each word and obtain their unified representations $\textbf{H}_u =\{\textbf{h}^u_i \}_{i=1}^M \in\mathcal{R}^{d_t\times M}$, where $\textbf{h}^u_i\in\mathcal{R}^{d_t}$ is the representation of the $i$-th word in $T^u$.
Besides, we can learn contextual word representations $\textbf{H}_c=\{\textbf{h}^c_i\}_{i=1}^M\in\mathcal{R}^{d_t\times M}$ for $T^c$ in the same way, where $\textbf{h}^c_i\in\mathcal{R}^{d_t}$ is the $i$-th word representation in $T^c$.

Finally, in general, different semantic aspects in clicked news and candidate usually have different importance for matching user interest with candidate news~\cite{wu2019hierarchical}.
For example, given a clicked news ``Apple's plans to make over-ear headphones.'', it contains two semantic aspects, i.e., ``Apple's product plan'' and ``headphones''.
The former is important for matching user interest with candidate news ``The best headphones of 2020.'' since users who are interested in headphones may click both of them.
While the latter is important for matching user interest with candidate news ``iPhone 12 cases buyer's guide.'' since users who are interested in the products of Apple may read them.
Thus, we apply a semantic co-attention network~\cite{shu2019defend,wu2020clickbait} to interactively learn semantic-based representations of news $n^u$ and $n^c$ by capturing semantic relatedness between their texts for interest matching.
Specifically, we first calculate the affinity matrix $\textbf{C}_t \in\mathcal{R}^{M\times M}$ measuring the semantic relevance between different words in texts $T^u$ and $T^c$:
\begin{equation}
    \textbf{C}_t =\textbf{H}_c^T \textbf{W}^t_c  \textbf{H}_u,
\end{equation}
where $\textbf{W}^t_c\in\mathcal{R}^{d_t\times d_t}$ is the trainable weights.
Then we compute the attention vector $\textbf{b}^u \in\mathcal{R}^M$ and $\textbf{b}^c \in\mathcal{R}^M$ for words in user's clicked news and candidate news respectively based on $\textbf{C}_t$:   
\begin{equation}
         \textbf{b}^u = \textbf{q}_t^T \cdot  \tanh( \textbf{W}^t_s \textbf{H}_u + \textbf{W}^t_h \textbf{H}_c f(\textbf{C}_t)),
\end{equation}
\begin{equation}
         \textbf{b}^c =  \textbf{q}_t^T \cdot  \tanh( \textbf{W}^t_s \textbf{H}_c + \textbf{W}^t_h \textbf{H}_u f(\textbf{C}^T_t)),
\end{equation}
where $\textbf{q}_t\in\mathcal{R}^{d_q} $ is the trainable attention query, $\textbf{W}^t_s\in\mathcal{R}^{d_q\times d_t}$ and $\textbf{W}^t_h \in\mathcal{R}^{d_q\times d_t}$ are trainable parameters.
Finally, we learn semantic-based representations $\textbf{t}^u \in\mathcal{R}^{d_t}$ and $\textbf{t}^c \in\mathcal{R}^{d_t}$ of news $n^u$ and $n^c$:
\begin{equation}
    \textbf{t}^u = \sum_{i=1}^M \beta^u_i \textbf{h}^u_i,  \quad  \quad    \beta^u_i = \frac{\exp(b^u_i)}{\sum_{j=1}^M\exp(b^u_j)},
\end{equation}
\begin{equation}
    \textbf{t}^c = \sum_{i=1}^M \beta^c_i \textbf{h}^c_i,  \quad  \quad    \beta^c_i = \frac{\exp(b^c_i)}{\sum_{j=1}^M\exp(b^c_j)},
\end{equation}
where $\beta^u_i$ and $\beta^c_i$ is weight of the $i$-th word in texts $T^u$ and $T^c$.

\subsection{User-News Co-Encoder}
We introduce our proposed \textit{user-news co-encoder}, which learns candidate news-aware user interest representation and user-aware candidate news representation from representations of user's clicked news and candidate news.
Usually, interests of a user are diverse, and only part of them can be matched with a candidate news~\cite{liu2019hi}.
Thus, learning candidate news-aware user interest representation can better model user interest for matching candidate news.
Similarly, a candidate news may cover multiple aspects, and a user may only be interested in part of them~\cite{wu2019ijcai,wu2019npa}.
Thus, learning user-aware candidate news representation is also beneficial for interest matching.
Thus, we apply a news co-attention network to learn candidate news-aware user representation and user-aware candidate news representation.
More specifically, we first calculate the affinity matrix $\textbf{C}_n \in\mathcal{R}^{N\times N}$ based on the representations of user's clicked news $\textbf{N}_u = \{\textbf{n}^u_i\}_{i=1}^N \in\mathcal{R}^{d_n\times N}$ and candidate news $\textbf{N}_c = \{\textbf{n}^c_i\}_{i=1}^N\in\mathcal{R}^{d_n\times N}$ to measure their relevance: 
\begin{equation}
    \textbf{C}_n = \textbf{N}_c^T \textbf{W}^n_c \textbf{N}_u,
\end{equation}
where $N$ denotes the number of clicked news, $\textbf{n}^u_i \in\mathcal{R}^{d_n}$ denotes the representation of user's $i$-th clicked news, $\textbf{n}^c_i\in\mathcal{R}^{d_n}$ denotes the corresponding representation of candidate news, and $\textbf{W}^n_c\in\mathcal{R}^{d_n\times d_n}$ is the trainable weights.
Then we compute the attention vector $\textbf{r}^u \in \mathcal{R}^{N} $ and $\textbf{r}^c \in \mathcal{R}^{N}$ for the representations of user's clicked news and candidate news based on the affinity matrix:
\begin{equation}
    \textbf{r}^u = \textbf{q}_n^T \cdot \tanh(\textbf{W}^n_s \textbf{N}_u +\textbf{W}^n_h \textbf{N}_c f(\textbf{C}_n) ),
\end{equation}
\begin{equation}
    \textbf{r}^c = \textbf{q}_n^T \cdot \tanh(\textbf{W}^n_s  \textbf{N}_c +\textbf{W}^n_h \textbf{N}_u f(\textbf{C}^T_n)),
\end{equation}
where $\textbf{q}_n\in\mathcal{R}^{d_q} $ denotes the trainable attention query, $\textbf{W}^n_s\in\mathcal{R}^{d_q\times d_n} $ and $\textbf{W}^n_h\in\mathcal{R}^{d_q\times d_n} $ are the trainable weights.
The candidate news-aware user representation $\textbf{u}\in\mathcal{R}^{ d_n} $ and user-aware candidate news representation $\textbf{c}\in\mathcal{R}^{ d_n}$ are formulated as:
\begin{equation}
    \textbf{u} = \sum_{i=1}^N \gamma^u_i \textbf{n}^u_i, \quad \quad \gamma^u_i = \frac{\exp(r^u_i)}{\sum_{j=1}^N \exp(r^u_j)},
\end{equation}
\begin{equation}
    \textbf{c} = \sum_{i=1}^N \gamma^c_i \textbf{n}^c_i, \quad \quad \gamma^c_i = \frac{\exp(r^c_i)}{\sum_{j=1}^N \exp(r^c_j)},
\end{equation}
where $\gamma^u_i$ and $\gamma^c_i$ denote attention weight of $\textbf{n}^u_i$ and $\textbf{n}^c_i$ respectively.

\subsection{Relevance Modeling and Model Training}
Following \citet{okura2017embedding}, we adopt dot product of candidate news-aware user representation $\textbf{u}$ and user-aware candidate news representation $\textbf{c}$ to measure the relevance $z\in\mathcal{R} $ of user interest and candidate news content, i.e., $z = \textbf{u}^T \cdot \textbf{c}$.
Candidate news are further recommended to the user based on their relevance scores.

Next, we introduce how we train the \textit{KIM} method.
We utilize the negative sampling technique~\cite{jiang2018recommendation,he2018adversarial} to construct the training dataset $\mathcal{S}$, where each positive sample is associated with $U$ negative sample randomly selected from the same news impression.
Then, we apply the NCE loss~\cite{oord2018representation} to formulate the loss function:
\begin{equation}
    \mathcal{L} = -\frac{1}{|\mathcal{S}|}\sum_{i=1}^{|\mathcal{S}|} \log( \frac{\exp(z^i_+)}{\exp(z^i_+)+\sum_{j=1}^U \exp(z^i_j)} ),
\end{equation}
where $\sigma$ denotes the sigmoid function, $z^i_+$ denotes the relevance score of the $i$-th positive sample, and $z^i_j$ denotes the relevance score of the $j$-th negative sample selected for the $i$-th positive sample.




Finally, we briefly discuss the computational complexity of \textit{KIM}.
Different from the methods that model user and candidate news independently, \textit{KIM} calculates representations of clicked news and candidate news collaboratively, which requires more computation resources because these representations cannot be prepared in advance.
Fortunately, in practice we can calculate contextual word embeddings $\textbf{H}$ and entity embeddings $\textbf{M}$ of different news offline and cache them to save the computational cost.

\section{Experiment}
\subsection{Datasets and Experimental Settings}

\begin{table}[h]
\centering
\caption{Detailed statistics of the \textit{MIND} and \textit{Feeds} datasets. }
\label{table.stat}
\resizebox{0.35\textwidth}{!}{\begin{tabular}{lcc}
\Xhline{1pt}
                               & \textit{MIND}    & \textit{Feeds}     \\ \hline
\# Users                       & 94,057  & 50,605    \\
\# Impressions                 & 230,117 & 210,000   \\
\# Clicks                      & 347,727 & 473,697   \\
\# News                        & 65,238  & 1,126,508 \\
Avg. \# words in news title    & 11.78   & 11.90     \\
Avg. \# entities in news title & 1.43    & 0.99      \\ 
Avg. \# neighbors in KG & 18.21 & 18.09      \\
\Xhline{1pt}
\end{tabular}
}
\end{table}

We evaluate the performance of different methods on the public \textit{MIND}\footnote{The small version of \textit{MIND} is used in our experiments.} dataset~\cite{wu2020mind} as well as a private dataset (named \textit{Feeds}) built from user logs of a commercial Feeds App in Microsoft.\footnote{We used the same techniques as in \textit{MIND} to protect user privacy in \textit{Feeds}.}
\textit{MIND} dataset was built from six-week sampled user logs in Microsoft News during Oct. 12 to Nov. 22, 2019, where the training and validation set were constructed by user logs in the fifth week, and the test set was constructed by user logs in the sixth week.
In \textit{MIND} dataset, entities in news titles were extracted and linked to WikiData automatically.
Their embeddings were trained based on the knowledge tuples extracted from WikiData via the TransE method~\cite{bordes2013translating}.
The \textit{Feeds} dataset was built from thirteen-week user logs during Jan. 23 to Apr. 01, 2020, where the training and validation set were constructed by 100,000 and 10,000 impressions randomly sampled from the first ten weeks respectively, and the test set was constructed by 100,000 impressions randomly sampled from the last three weeks.
Following~\citet{wu2020mind}, in \textit{Feeds} dataset we also extracted entities in news titles and pre-trained their embeddings based on the WikiData.
In these two datasets, we used news titles as news texts.
Besides, we used WikiData as the knowledge graph in experiments.
More detailed statistics is listed in Table~\ref{table.stat}.

	   



\begin{table*}[]

\caption{Results of different methods on the two  datasets. We perform t-test on these results which shows \textit{KIM} can significantly (at the level p < 0.01) outperform all baseline methods.}
\label{table.performance}

\centering
\resizebox{0.95\textwidth}{!}{
\begin{tabular}{ccccccccc}
\Xhline{1.2pt}
      & \multicolumn{4}{c}{ \textit{MIND} }                                                               & \multicolumn{4}{c}{\textit{Feeds}  }                                         \\ \hline
      & AUC            & MRR            & nDCG@5         & nDCG@10                             & AUC            & MRR            & nDCG@5         & nDCG@10        \\ \hline
\multicolumn{1}{c|}{EBNR} & 61.28$\pm$0.27 & 27.77$\pm$0.21 & 30.10$\pm$0.28 & \multicolumn{1}{c|}{36.75$\pm$0.24} & 63.44$\pm$0.39 & 27.97$\pm$0.25 & 32.01$\pm$0.32 & 37.57$\pm$0.35 \\
\multicolumn{1}{c|}{DKN}   & 64.08$\pm$0.12 & 29.06$\pm$0.16 & 31.82$\pm$0.11 & \multicolumn{1}{c|}{38.52$\pm$0.14} & 62.91$\pm$0.26 & 28.08$\pm$0.20 & 32.20$\pm$0.24 & 37.75$\pm$0.22 \\
\multicolumn{1}{c|}{DAN}   & 65.14$\pm$0.16 & 30.04$\pm$0.20 & 32.98$\pm$0.22 & \multicolumn{1}{c|}{39.52$\pm$0.19} & 62.65$\pm$0.49 & 27.79$\pm$0.32 & 31.79$\pm$0.40 & 37.37$\pm$0.39 \\
\multicolumn{1}{c|}{NAML}  & 64.21$\pm$0.20 & 29.71$\pm$0.13 & 32.51$\pm$0.20 & \multicolumn{1}{c|}{39.00$\pm$0.12} & 64.24$\pm$0.38 & 28.81$\pm$0.21 & 33.06$\pm$0.28 & 38.52$\pm$0.29 \\
\multicolumn{1}{c|}{NPA}   & 63.71$\pm$0.27 & 29.84$\pm$0.12 & 32.40$\pm$0.19 & \multicolumn{1}{c|}{39.02$\pm$0.20} & 63.69$\pm$0.75 & 28.51$\pm$0.47 & 32.74$\pm$0.64 & 38.27$\pm$0.62 \\
\multicolumn{1}{c|}{LSTUR} & 65.51$\pm$0.29 & 30.22$\pm$0.31 & 33.26$\pm$0.38 & \multicolumn{1}{c|}{39.76$\pm$0.34} & 64.66$\pm$0.33 & 29.04$\pm$0.26 & 33.44$\pm$0.32 & 38.82$\pm$0.30 \\
\multicolumn{1}{c|}{NRMS}  & 65.36$\pm$0.21 & 30.02$\pm$0.11 & 33.11$\pm$0.15 & \multicolumn{1}{c|}{39.61$\pm$0.14} & 65.15$\pm$0.13 & 29.29$\pm$0.12 & 33.78$\pm$0.13 & 39.24$\pm$0.13 \\
\multicolumn{1}{c|}{FIM}   & 64.46$\pm$0.22 & 29.52$\pm$0.26 & 32.26$\pm$0.24 & \multicolumn{1}{c|}{39.08$\pm$0.27} & 65.67$\pm$0.20 & 29.83$\pm$0.24 & 34.51$\pm$0.31 & 39.97$\pm$0.25 \\
\multicolumn{1}{c|}{KRED}& 65.61$\pm$0.35 & 30.63$\pm$0.27 & 33.80$\pm$0.24 & \multicolumn{1}{c|}{40.23$\pm$0.27} & 65.47$\pm$0.07 & 29.59$\pm$0.04 & 34.15$\pm$0.05 & 39.69$\pm$0.05 \\ \hline
\multicolumn{1}{c|}{KIM}  & \textbf{67.13}$\pm$0.29 & \textbf{32.08}$\pm$0.24 & \textbf{35.49}$\pm$0.34 & \multicolumn{1}{c|}{\textbf{41.79}$\pm$0.28} & \textbf{66.45}$\pm$0.13 & \textbf{30.27}$\pm$0.09 & \textbf{35.04}$\pm$0.09 & \textbf{40.43}$\pm$0.12 \\ 
\Xhline{1.2pt}
\end{tabular}
}
\end{table*}

Next, we introduce hyper-parameters of \textit{KIM}\footnote{Codes are available at https://github.com/JulySinceAndrew/KIM-SIGIR-2021.} and experiment settings.
For each news, we only used the first 30 words in news titles and first 5 entities in news.
We randomly sampled 10 neighbors for each entity from the knowledge graph.
Besides, we only used the recent 50 clicked news of each user.
The word and entity embedding vectors were initialized by 300-dimensional glove embeddings~\cite{pennington2014glove} and 100-dimensional TransE embeddings~\cite{bordes2013translating}, respectively.
Due to limitation of GPU memory, we only fine-tuned word embeddings and did not fine-tune entity embeddings in experiments.
In \textit{semantic co-encoder}, the transformer contained 10 attention heads and output vectors of each head were 40-dimensional.
Besides, the CNN network contained 400 filters.
In \textit{knowledge co-encoder}, all multi-head self-attention networks in the graph attention and co-attention networks contained 5 attention heads, and all of these heads output 20-dimensional vectors.
Besides, all attention queries in \textit{KIM} were set to 100-dimension.
For effective model training we applied the dropout technique~\cite{srivastava2014dropout} with 0.2 dropout probability.
We sampled $4$ negative samples for each positive sample.
We utilized Adam optimizer~\cite{kingma2014adam} to train \textit{KIM} with $5\times10^{-5}$ learning rate.
All hyper-parameters of \textit{KIM} and other baseline methods were selected based on the validation dataset.
Following previous works~\cite{wu2019ijcai}, we used AUC, MRR, nDCG\@5, and nDCG\@10 for evaluation.

\subsection{Performance Evaluation}

We compare \textit{KIM} with several state-of-the-art personalized news recommendation methods, which are listed as follow:
(1) \textit{EBNR}~\cite{okura2017embedding}: representing user interest from user's click history via a GRU network.
(2) \textit{DKN}~\cite{wang2018dkn}: applying a multi-channel CNN network~\cite{lecun1998gradient} to embeddings of aligned words and entities in news titles to learn news representations.
(3) \textit{DAN}~\cite{danzhu2019}: learning news representations from words and entities of news titles via a CNN network, and learning user interest representations via an attentive LSTM network~\cite{hochreiter1997long}.
(4) \textit{NAML}~\cite{wu2019ijcai}: learning news representations from news titles, bodies, categories, and sub-categories via multiple attentive CNN networks.
(5) \textit{NPA}~\cite{wu2019npa}: using attention networks with personalized attention queries to learn news and user representations.
(6) \textit{LSTUR}~\cite{an2019neural}: modeling shot-term user interests from user's recent clicked news via a GRU network and modeling long-term user interest via user ID embeddings.
(7) \textit{NRMS}~\cite{wu2019neuralc}: modeling news content and user click behaviors via multi-head self-attention networks.
(8) \textit{KRED}~\cite{liu2020kred}: learning representations for news from the entities in news and their neighbors in the knowledge graph via a graph attention network.
(9) \textit{FIM}~\cite{wang2020fine}: matching user and news from texts of users' clicked news and candidate news via CNN networks.

We repeat different experiments five times and list the average performance of different methods and corresponding standard deviations in Table~\ref{table.performance}.
First, we can find \textit{KIM} significantly outperforms other baseline methods which independently model candidate news and user interest without consideration of their relatedness, \textit{LSTUR}, \textit{NRMS} and \textit{KRED}.
This is because a user may be interested in multiple areas, and a candidate news may also contain multiple aspects and entities.
Thus, it is difficult for these methods to accurately match user interest and candidate news since they are independently modeled in these methods.
Different from these methods, in our \textit{KIM} method we propose a knowledge-aware interactive matching framework to interactively model user interest and candidate news.
Our \textit{KIM} can effectively incorporate relatedness between clicked news and candidate news at both semantic and knowledge levels for better interest matching.
Second, \textit{KIM} also outperforms baseline methods which model user interest with the consideration of candidate news, such as \textit{DKN}, \textit{DAN}.
This is because candidate news may cover multiple aspects, and a user may only be interested in a part of them~\cite{wu2019ijcai,wu2019npa}.
However, these methods model candidate news without the consideration of the target user, which may be inferior for further matching candidate news with user interest.
Different from these methods, our \textit{KIM} can model candidate news with target user information.
Besides, in these methods clicked news and candidate news are also independently modeled from their content without consideration of their relatedness, which may be suboptimal for further measuring the relevance with candidate news and user interest inferred from clicked news.
Different from these methods, in our \textit{KIM} we propose a \textit{knowledge co-encoder} and a \textit{semantic co-encoder} to interactively learn knowledge-aware representations of both clicked news and candidate news.


\subsection{Ablation Study}

\begin{figure*}[htbp]
\begin{minipage}[t]{0.31\textwidth}
\resizebox{\textwidth}{!}{\includegraphics{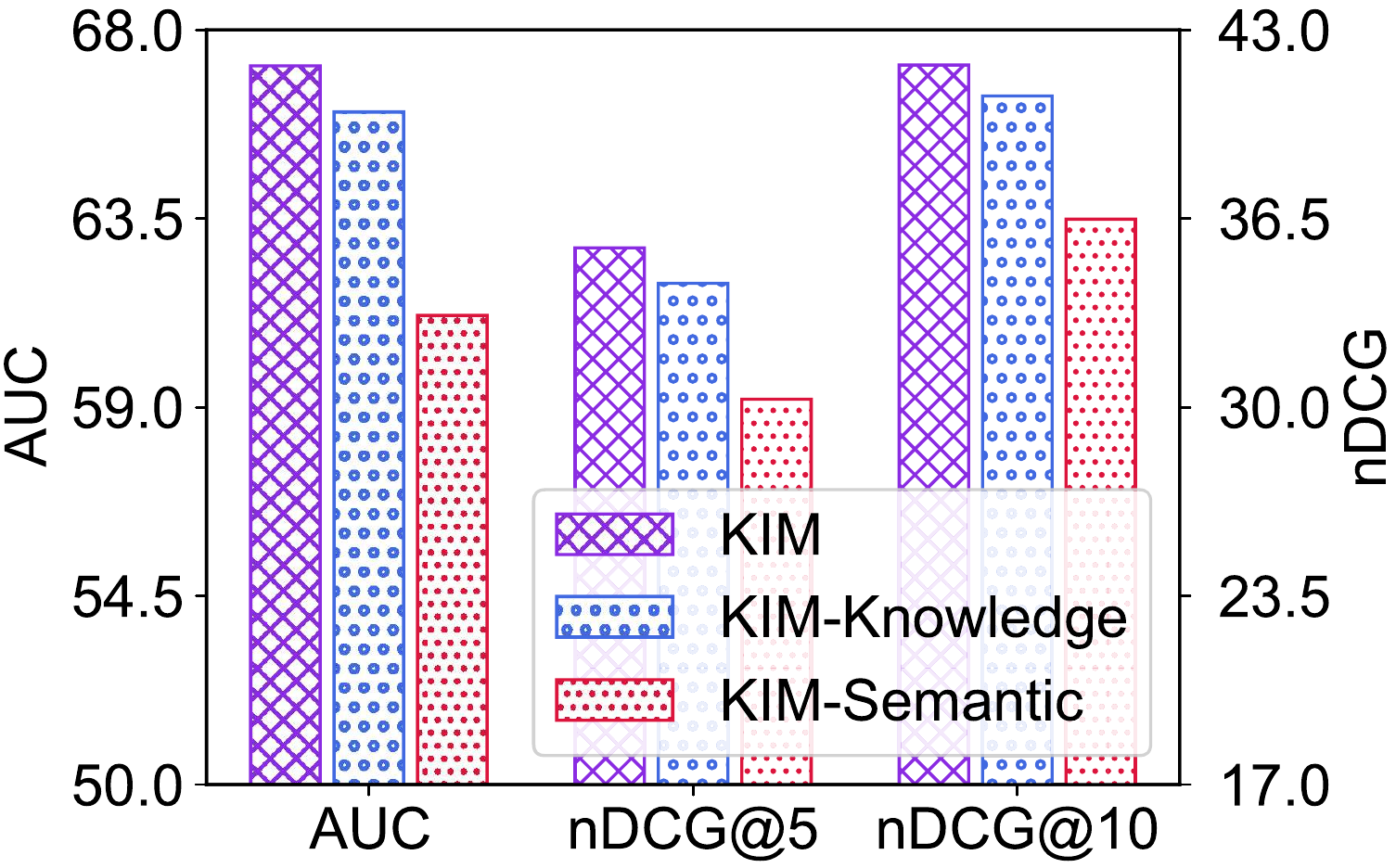}}
\caption{\textit{KIM} with different news information removed.}
\label{fig.information}
\end{minipage} 
\hfill
\begin{minipage}[t]{0.31\textwidth}
\resizebox{\textwidth}{!}{\includegraphics{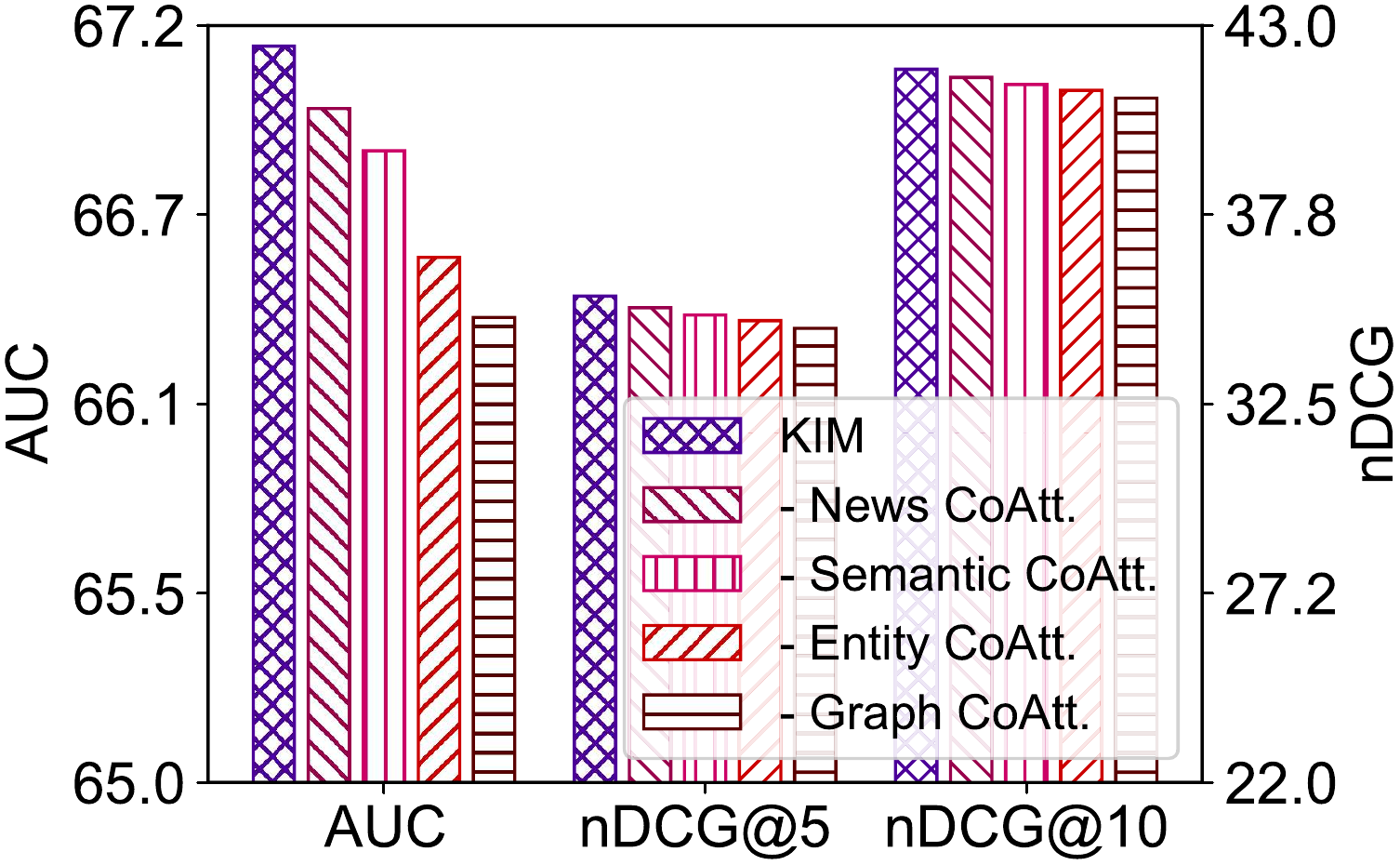}}
    \caption{\textit{KIM} with different module removed. CoAtt. is  co-attention.}
    \label{fig.ablation}
\end{minipage}
\hfill
\begin{minipage}[t]{0.31\textwidth}
\resizebox{\textwidth}{!}{\includegraphics{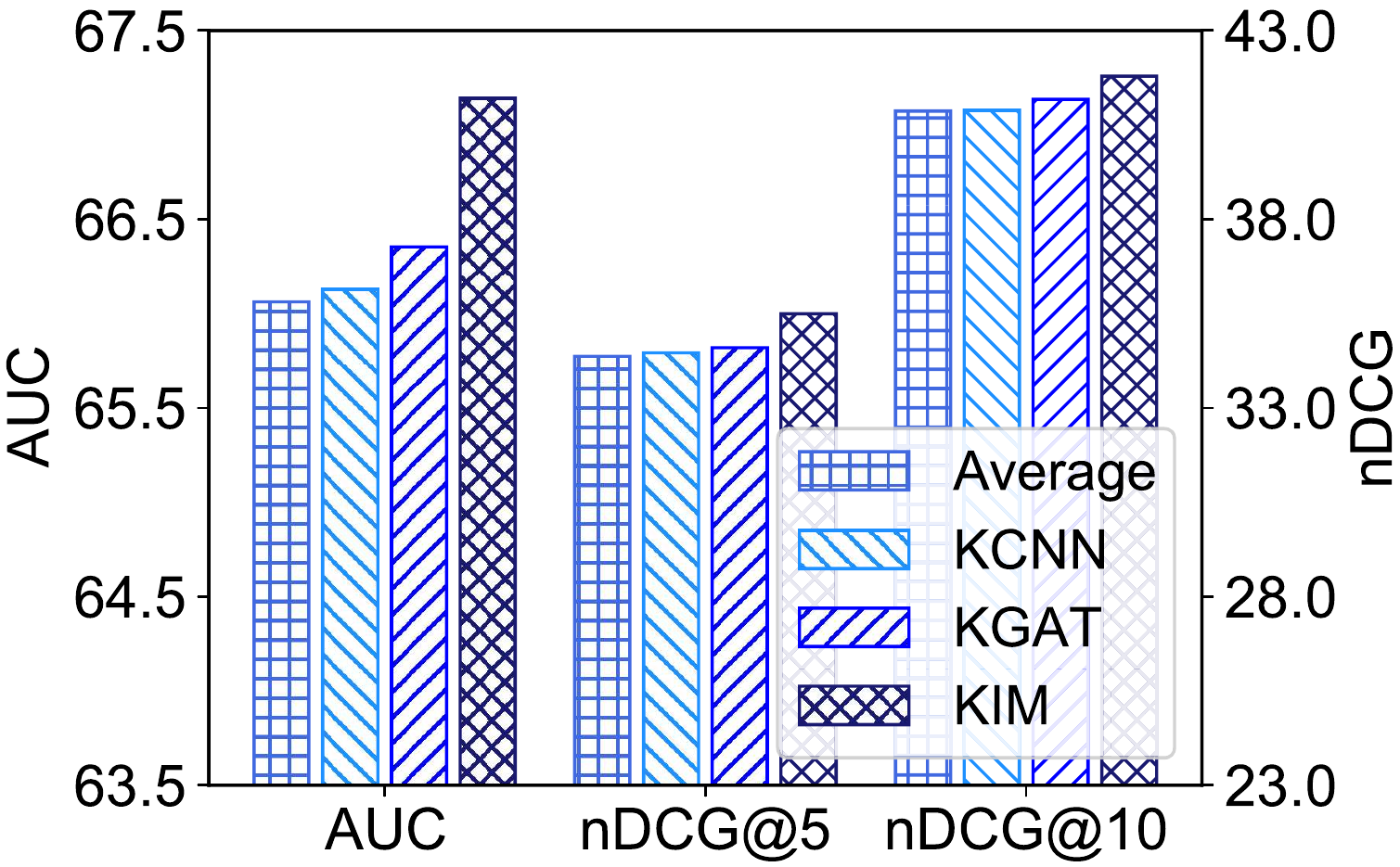}}
    \caption{\textit{KIM} with different knowledge modeling methods.}
    \label{fig.knowledge}
\end{minipage}
\end{figure*}

In this section, we conduct two ablation studies to evaluate the effectiveness of \textit{KIM}.
We first evaluate the effectiveness of different information, i.e., texts and knowledge, for news content modeling.
Due to space limitation, we only show the experimental results on \textit{MIND} in the following sections.
The experimental results are shown in Fig.~\ref{fig.information}, from which we have several observations.
First, removing semantic information (i.e., news texts) seriously hurts the performance of \textit{KIM}.
This is because texts usually contain rich information on news content and are vitally important for news content understanding~\cite{wu2020mind}.
Removing semantic information makes the news representations lose much important information and cannot model news content accurately.
Second, removing knowledge (i.e., entities and their neighbors in the knowledge graph) in news content modeling also makes the performance of \textit{KIM} decline significantly.
This is because textual information is usually insufficient to understand news content~\cite{wang2018dkn,liu2020kred}.
Fortunately, knowledge graph contains rich relatedness between different entities.
Moreover, relatedness between entities in user's clicked news and candidate news can provide rich information beyond semantic information for understanding user interest in candidate news.
Thus, incorporating entity information into personalized news recommendation has the potential to improve the accuracy of recommendation.




Next, we evaluate the effectiveness of several important co-attention networks in \textit{KIM} by replacing them with attention networks individually.
Fig.~\ref{fig.ablation} shows the experimental results, from which we have several findings.
First, after removing the news co-attention network in \textit{user-news co-encoder}, the performance of \textit{KIM} gets worse.
This is because user interest may be diverse, and only a part of user's clicked news is informative for modeling the relevance between user interest and candidate news~\cite{wang2018dkn}.
Besides, candidate news content may contain multiple aspects and a user may be interested in only a part of them.
Thus, learning candidate news-aware user interest and user-aware candidate news representation via a news co-attention network can better capture user interest in candidate news.
Second, removing the semantic co-attention network also hurts the performance of \textit{KIM}.
This is because semantic relatedness between clicked news and candidate news can help understand user interest in candidate news.
Besides, a candidate news or a clicked news usually contains multiple aspects, and only a part of them is useful for the interest matching.
It is difficult to effectively capture the relatedness of clicked news and candidate news at semantic level if their semantic information are independently modeled.
Thus, interactively learning semantic-based representations of clicked news and candidate news via a semantic co-attention network can better capture relatedness between them for matching user interest with candidate news.
Third, removing both the graph co-attention network and entity co-attention network makes the performance of \textit{KIM} decline.
This is because relatedness between clicked news and candidate news at entity level is also very informative for interest matching.
Besides, it is also suboptimal for interest matching if the method represents clicked news and candidate news from their entities independently.
In \textit{KIM} method, both the graph co-attention network and entity co-attention network are used to capture relatedness between entities of clicked news and candidate news in an interactive way, which can incorporate rich information into \textit{KIM} model for interest matching.

\subsection{Effectiveness of Knowledge Modeling}


We evaluate the effectiveness of the \textit{knowledge co-encoder} in \textit{KIM} by comparing \textit{KIM} with its variations which independently model clicked and candidate news from their entities.
The first one is \textit{Average}, which averages embeddings of entities in news and their neighbors within $K$ hops as the knowledge-based news representations.
The second one is \textit{KCNN}, which learns knowledge-based news representations from entities and their neighbors via the KCNN network proposed in \textit{DKN}~\cite{wang2018dkn}.
The third one is \textit{KGAT}, which uses a knowledge graph attention network proposed in \textit{KRED}~\cite{liu2020kred} to learn knowledge-based news representations from entities in news and their neighbors on the knowledge graph.
Besides, all of these variations have the same text modeling method with \textit{KIM} for fair comparisons.
Fig.~\ref{fig.knowledge} shows the experimental results.

First, \textit{Average} has the worst performance among these methods.
This is because different entities in news and their neighbors usually have different informativeness for news content understanding.
Since \textit{Average} ignores the relative importance of different entities, it cannot effectively model news content based on entities.
Second, \textit{KGAT} outperforms \textit{KCNN}.
This is because there is usually conceptual relatedness between different neighbors of an entity.
\textit{KCNN} only uses the average embeddings of neighbors of entities in news to enhance their representations and ignores such relatedness.
Different from \textit{DKN}, \textit{KGAT} utilizes a graph attention network to model the relatedness between neighbor entities, which can learn more accurate entity representations.
Third, \textit{KIM} significantly outperforms all of baseline methods, i.e., \textit{Avg}, \textit{KCNN}, \textit{KGAT}.
This is because relatedness between clicked news and candidate news at entity level can provide rich clues to infer user interests in candidate news.
Besides, a clicked news or a candidate news may contain multiple entities and not all of them are useful for matching user interest with candidate news.
However, these methods independently model entity information for clicked news and candidate news without consideration of their relatedness, which is suboptimal for further matching candidate news with user interest inferred from click history.
Different from these methods, we propose a \textit{knowledge co-encoder} to interactively learn knowledge-based representations for clicked news and candidate news from the relatedness between their entities for better interest matching.

\subsection{Influence of Hyper-parameters}
\begin{figure}
    \centering
    \resizebox{0.32\textwidth}{!}{
    \includegraphics{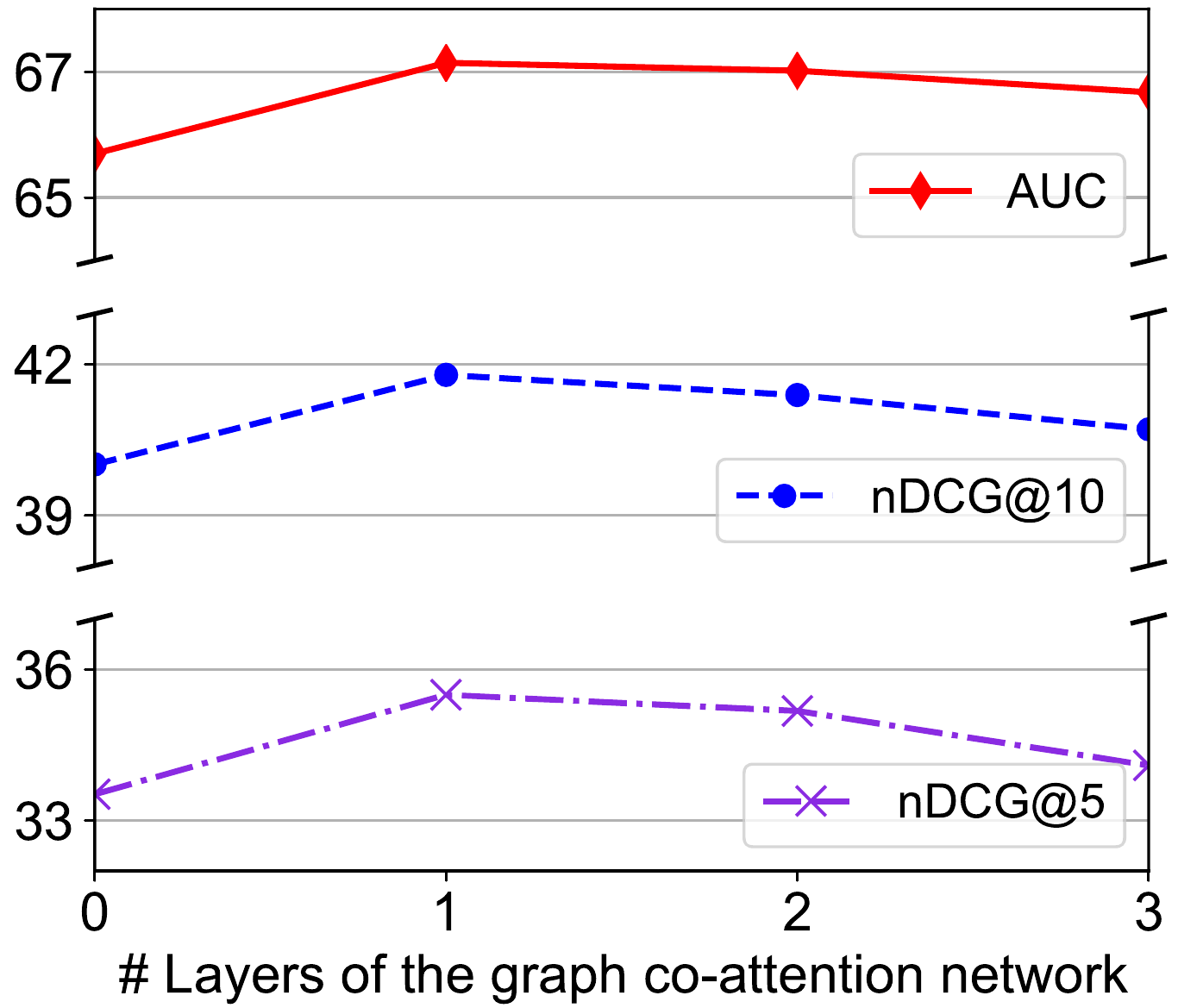}
    }
    \caption{Performance of \textit{KIM} under different number of layers in the graph co-attention network, i.e., $K$.}
    \label{fig.hyper}
\end{figure}

We evaluate the influence of an important hyper-parameter, i.e., the number of layers of the graph co-attention network, i.e., $K$, on the performance of \textit{KIM}.
Results are shown in Fig.~\ref{fig.hyper}, from which we have two observations.
First, the performance of \textit{KIM} first increases with the increase of $K$.
This is because the relatedness between entities in clicked news and candidate news is informative for understanding user interest in candidate news.
Besides, the GCAT network stacked for $K$ layers can incorporate neighbors of entities in clicked news and candidate news within $K$ hops for learning their representations.
When $K$ is too small, the relatedness between user's clicked news and candidate news cannot be fully explored based on their entities, which is harmful to the recommendation accuracy.
Second, when $K$ is too large, the performance of \textit{KIM} begins to decline.
This is because when $K$ becomes too large, too many multi-hop neighbors are considered when modeling the relatedness between user's clicked news and candidate news.
This may bring much noise to the \textit{KIM} model and hurt the recommendation accuracy.
Thus, a moderate value of $K$, i.e., 1, is suitable for \textit{KIM}.

\subsection{Case Study}

\begin{figure}
    \centering
    \resizebox{0.45\textwidth}{!}{\includegraphics{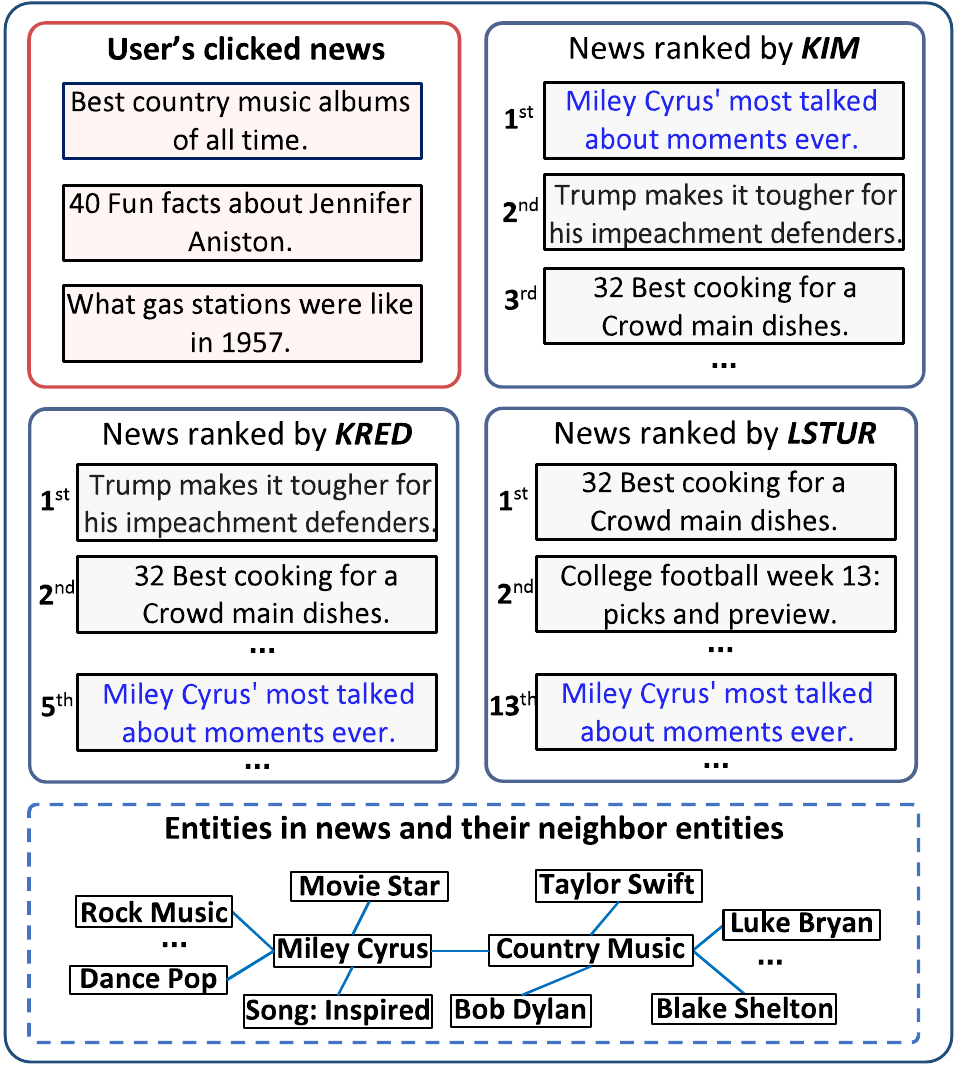}}
    \caption{News recommended to a randomly selected user by different methods. The news in blue is the news actually clicked by this user in this impression. The historical clicked news of this user are also shown in this figure.}
    
    \label{fig.case}

\end{figure}

We conduct a case study to show the effectiveness of \textit{KIM} by comparing it with \textit{LSTUR} and \textit{KRED}.
We compare \textit{LSTUR} since it achieves the best performance (Table~\ref{table.performance}) among baseline methods which model news content from pain news texts.
Besides we compare \textit{KRED} since it achieves the best performance (Table~\ref{table.performance}) among knowledge-aware baseline methods.
We show the reading history of a randomly sampled user, and the news recommended by these methods in the same impression where the user only clicked one candidate news in Fig.~\ref{fig.case}, from which we have several observations.
First, both \textit{KRED} and \textit{KIM} rank the candidate news clicked by the user higher than \textit{LSTUR}.
This is because it is difficult to understand the relevance of user interest and candidate news from the textual information of user's clicked news and candidate news.
However, since Miley Cryus is a representative singer of country music, on the knowledge graph we can find that the entity ``Country Music'' in the first clicked news of the user has a link with the entity ``Miley Cryus'' in the candidate news clicked by the user.
Thus, based on the information provided by the knowledge graph, \textit{KRED} and \textit{KIM} can better understand the relevance of user interest and candidate news.
Second, \textit{KIM} ranks the candidate news clicked by the user higher than \textit{KRED}.
This is because both of these two entities have rich relatedness with many other neighbor entities on the knowledge graph.
For example, besides ``Miley Cyrus'', the entity ``Country Music'' also has relatedness with many other representative singers such as ``Bob Dylan'', ``Talyor Swift'', and so on.
In addition, the entity ``Miley Cryus'' also has relatedness with the entities of other areas which ``Miley Cryus'' is skilled in, such as ``rock music'', ``dance-pop'' and so on.
However, it is difficult for \textit{KRED} which independently model user's clicked news and candidate news to accurately capture the useful relatedness between entities of clicked news and candidate news for interest matching.
Different from \textit{KRED}, \textit{KIM} uses a \textit{knowledge co-encoder} to interactively represent clicked news and candidate news from their relatedness at entity level, which can better capture user interest in candidate news than \textit{KRED}.

\section{CONCLUSION}

In this paper, we propose a knowledge-aware interactive matching framework for personalized news recommendation (named \textit{KIM}).
The framework aims to interactively model candidate news and user interests for more accurate interest matching.
More specifically, we first propose a graph co-attention network to model entities based on the knowledge graph by selecting and aggregating the information of their neighbors which are informative for interest matching.
We also propose to use an entity co-attention network to interactively model clicked news and candidate news from relatedness between their entities.
Besides, we propose to use a semantic co-attention network to interactively model clicked news and candidate news from semantic relatedness between their texts.
Moreover, we propose a \textit{user-news co-encoder} to learn candidate news-aware user representation and user-aware candidate news representation to better capture the relevance between user interest and candidate news.
We conduct extensive experiments on two real-world datasets.
The experimental results show that our \textit{KIM} method can significantly outperform other baseline methods.

\section*{Acknowledgments}
This work was supported by the National Natural Science Foundation of China under Grant numbers U1936208, U1936216, 61862002, and U1705261.
We are grateful to Xing Xie, Tao Di, Wei He, Andy Jing, and Jie Li in Microsoft for their great comments and suggestions on this work.

\bibliographystyle{ACM-Reference-Format}
\bibliography{mybib}


\begin{thebibliography}{47}


\ifx \showCODEN    \undefined \def \showCODEN     #1{\unskip}     \fi
\ifx \showDOI      \undefined \def \showDOI       #1{#1}\fi
\ifx \showISBNx    \undefined \def \showISBNx     #1{\unskip}     \fi
\ifx \showISBNxiii \undefined \def \showISBNxiii  #1{\unskip}     \fi
\ifx \showISSN     \undefined \def \showISSN      #1{\unskip}     \fi
\ifx \showLCCN     \undefined \def \showLCCN      #1{\unskip}     \fi
\ifx \shownote     \undefined \def \shownote      #1{#1}          \fi
\ifx \showarticletitle \undefined \def \showarticletitle #1{#1}   \fi
\ifx \showURL      \undefined \def \showURL       {\relax}        \fi
\providecommand\bibfield[2]{#2}
\providecommand\bibinfo[2]{#2}
\providecommand\natexlab[1]{#1}
\providecommand\showeprint[2][]{arXiv:#2}

\bibitem[\protect\citeauthoryear{An, Wu, Wu, Zhang, Liu, and Xie}{An
  et~al\mbox{.}}{2019}]%
        {an2019neural}
\bibfield{author}{\bibinfo{person}{Mingxiao An}, \bibinfo{person}{Fangzhao Wu},
  \bibinfo{person}{Chuhan Wu}, \bibinfo{person}{Kun Zhang},
  \bibinfo{person}{Zheng Liu}, {and} \bibinfo{person}{Xing Xie}.}
  \bibinfo{year}{2019}\natexlab{}.
\newblock \showarticletitle{Neural news recommendation with long-and short-term
  user representations}. In \bibinfo{booktitle}{\emph{ACL}}.
  \bibinfo{pages}{336--345}.
\newblock


\bibitem[\protect\citeauthoryear{Bansal, Das, and Bhattacharyya}{Bansal
  et~al\mbox{.}}{2015}]%
        {bansal2015content}
\bibfield{author}{\bibinfo{person}{Trapit Bansal}, \bibinfo{person}{Mrinal
  Das}, {and} \bibinfo{person}{Chiranjib Bhattacharyya}.}
  \bibinfo{year}{2015}\natexlab{}.
\newblock \showarticletitle{Content driven user profiling for comment-worthy
  recommendations of news and blog articles}. In
  \bibinfo{booktitle}{\emph{RecSys.}} \bibinfo{pages}{195--202}.
\newblock


\bibitem[\protect\citeauthoryear{Bordes, Usunier, Garcia-Duran, Weston, and
  Yakhnenko}{Bordes et~al\mbox{.}}{2013}]%
        {bordes2013translating}
\bibfield{author}{\bibinfo{person}{Antoine Bordes}, \bibinfo{person}{Nicolas
  Usunier}, \bibinfo{person}{Alberto Garcia-Duran}, \bibinfo{person}{Jason
  Weston}, {and} \bibinfo{person}{Oksana Yakhnenko}.}
  \bibinfo{year}{2013}\natexlab{}.
\newblock \showarticletitle{Translating embeddings for modeling
  multi-relational data}. In \bibinfo{booktitle}{\emph{NIPS}}.
  \bibinfo{pages}{2787--2795}.
\newblock


\bibitem[\protect\citeauthoryear{Das, Datar, Garg, and Rajaram}{Das
  et~al\mbox{.}}{2007}]%
        {das2007google}
\bibfield{author}{\bibinfo{person}{Abhinandan~S Das}, \bibinfo{person}{Mayur
  Datar}, \bibinfo{person}{Ashutosh Garg}, {and} \bibinfo{person}{Shyam
  Rajaram}.} \bibinfo{year}{2007}\natexlab{}.
\newblock \showarticletitle{Google news personalization: scalable online
  collaborative filtering}. In \bibinfo{booktitle}{\emph{WWW}}.
  \bibinfo{pages}{271--280}.
\newblock


\bibitem[\protect\citeauthoryear{Ge, Wu, Wu, Qi, and Huang}{Ge
  et~al\mbox{.}}{2020}]%
        {ge2020graph}
\bibfield{author}{\bibinfo{person}{Suyu Ge}, \bibinfo{person}{Chuhan Wu},
  \bibinfo{person}{Fangzhao Wu}, \bibinfo{person}{Tao Qi}, {and}
  \bibinfo{person}{Yongfeng Huang}.} \bibinfo{year}{2020}\natexlab{}.
\newblock \showarticletitle{Graph enhanced representation learning for news
  recommendation}. In \bibinfo{booktitle}{\emph{WWW}}.
  \bibinfo{pages}{2863--2869}.
\newblock


\bibitem[\protect\citeauthoryear{Hamaguchi, Oiwa, Shimbo, and
  Matsumoto}{Hamaguchi et~al\mbox{.}}{2017}]%
        {hamaguchi2017knowledge}
\bibfield{author}{\bibinfo{person}{Takuo Hamaguchi}, \bibinfo{person}{Hidekazu
  Oiwa}, \bibinfo{person}{Masashi Shimbo}, {and} \bibinfo{person}{Yuji
  Matsumoto}.} \bibinfo{year}{2017}\natexlab{}.
\newblock \showarticletitle{Knowledge transfer for out-of-knowledge-base
  entities: a graph neural network approach}. In
  \bibinfo{booktitle}{\emph{IJCAI}}. \bibinfo{pages}{1802--1808}.
\newblock


\bibitem[\protect\citeauthoryear{He, He, Du, and Chua}{He
  et~al\mbox{.}}{2018}]%
        {he2018adversarial}
\bibfield{author}{\bibinfo{person}{Xiangnan He}, \bibinfo{person}{Zhankui He},
  \bibinfo{person}{Xiaoyu Du}, {and} \bibinfo{person}{Tat-Seng Chua}.}
  \bibinfo{year}{2018}\natexlab{}.
\newblock \showarticletitle{Adversarial personalized ranking for
  recommendation}. In \bibinfo{booktitle}{\emph{SIGIR}}.
  \bibinfo{pages}{355--364}.
\newblock


\bibitem[\protect\citeauthoryear{Hochreiter and Schmidhuber}{Hochreiter and
  Schmidhuber}{1997}]%
        {hochreiter1997long}
\bibfield{author}{\bibinfo{person}{Sepp Hochreiter} {and}
  \bibinfo{person}{J{\"u}rgen Schmidhuber}.} \bibinfo{year}{1997}\natexlab{}.
\newblock \showarticletitle{Long short-term memory}.
\newblock \bibinfo{journal}{\emph{Neural computation}} (\bibinfo{year}{1997}),
  \bibinfo{pages}{1735--1780}.
\newblock


\bibitem[\protect\citeauthoryear{Jiang, Liu, Fu, Wu, and Zhang}{Jiang
  et~al\mbox{.}}{2018}]%
        {jiang2018recommendation}
\bibfield{author}{\bibinfo{person}{Zhengshen Jiang}, \bibinfo{person}{Hongzhi
  Liu}, \bibinfo{person}{Bin Fu}, \bibinfo{person}{Zhonghai Wu}, {and}
  \bibinfo{person}{Tao Zhang}.} \bibinfo{year}{2018}\natexlab{}.
\newblock \showarticletitle{Recommendation in heterogeneous information
  networks based on generalized random walk model and bayesian personalized
  ranking}. In \bibinfo{booktitle}{\emph{WSDM}}. \bibinfo{pages}{288--296}.
\newblock


\bibitem[\protect\citeauthoryear{Khattar, Kumar, Varma, and Gupta}{Khattar
  et~al\mbox{.}}{2018}]%
        {khattar2018weave}
\bibfield{author}{\bibinfo{person}{Dhruv Khattar}, \bibinfo{person}{Vaibhav
  Kumar}, \bibinfo{person}{Vasudeva Varma}, {and} \bibinfo{person}{Manish
  Gupta}.} \bibinfo{year}{2018}\natexlab{}.
\newblock \showarticletitle{Weave\&Rec: A word embedding based 3-D
  Convolutional network for news recommendation}. In
  \bibinfo{booktitle}{\emph{CIKM}}. \bibinfo{pages}{1855--1858}.
\newblock


\bibitem[\protect\citeauthoryear{Kim}{Kim}{2014}]%
        {kim2014convolutional}
\bibfield{author}{\bibinfo{person}{Yoon Kim}.} \bibinfo{year}{2014}\natexlab{}.
\newblock \showarticletitle{Convolutional neural networks for sentence
  classification}. In \bibinfo{booktitle}{\emph{EMNLP}}.
  \bibinfo{pages}{1746--1751}.
\newblock


\bibitem[\protect\citeauthoryear{Kingma and Ba}{Kingma and Ba}{2015}]%
        {kingma2014adam}
\bibfield{author}{\bibinfo{person}{Diederik~P Kingma} {and}
  \bibinfo{person}{Jimmy Ba}.} \bibinfo{year}{2015}\natexlab{}.
\newblock \showarticletitle{Adam: A Method for Stochastic Optimization}. In
  \bibinfo{booktitle}{\emph{ICLR}}.
\newblock


\bibitem[\protect\citeauthoryear{Kompan and Bielikov{\'a}}{Kompan and
  Bielikov{\'a}}{2010}]%
        {kompan2010content}
\bibfield{author}{\bibinfo{person}{Michal Kompan} {and}
  \bibinfo{person}{M{\'a}ria Bielikov{\'a}}.} \bibinfo{year}{2010}\natexlab{}.
\newblock \showarticletitle{Content-based news recommendation}. In
  \bibinfo{booktitle}{\emph{EC-Web}}. \bibinfo{pages}{61--72}.
\newblock


\bibitem[\protect\citeauthoryear{Konstan, Miller, Maltz, Herlocker, Gordon, and
  Riedl}{Konstan et~al\mbox{.}}{1997}]%
        {konstan1997grouplens}
\bibfield{author}{\bibinfo{person}{Joseph~A Konstan},
  \bibinfo{person}{Bradley~N Miller}, \bibinfo{person}{David Maltz},
  \bibinfo{person}{Jonathan~L Herlocker}, \bibinfo{person}{Lee~R Gordon}, {and}
  \bibinfo{person}{John Riedl}.} \bibinfo{year}{1997}\natexlab{}.
\newblock \showarticletitle{GroupLens: applying collaborative filtering to
  Usenet news}.
\newblock \bibinfo{journal}{\emph{Commun. ACM}} (\bibinfo{year}{1997}),
  \bibinfo{pages}{77--87}.
\newblock


\bibitem[\protect\citeauthoryear{LeCun, Bottou, Bengio, Haffner,
  et~al\mbox{.}}{LeCun et~al\mbox{.}}{1998}]%
        {lecun1998gradient}
\bibfield{author}{\bibinfo{person}{Yann LeCun}, \bibinfo{person}{L{\'e}on
  Bottou}, \bibinfo{person}{Yoshua Bengio}, \bibinfo{person}{Patrick Haffner},
  {et~al\mbox{.}}} \bibinfo{year}{1998}\natexlab{}.
\newblock \showarticletitle{Gradient-based learning applied to document
  recognition}.
\newblock \bibinfo{journal}{\emph{Proc. IEEE}} (\bibinfo{year}{1998}),
  \bibinfo{pages}{2278--2324}.
\newblock


\bibitem[\protect\citeauthoryear{Lian, Zhang, Xie, and Sun}{Lian
  et~al\mbox{.}}{2018}]%
        {lian2018towards}
\bibfield{author}{\bibinfo{person}{Jianxun Lian}, \bibinfo{person}{Fuzheng
  Zhang}, \bibinfo{person}{Xing Xie}, {and} \bibinfo{person}{Guangzhong Sun}.}
  \bibinfo{year}{2018}\natexlab{}.
\newblock \showarticletitle{Towards better representation learning for
  personalized news recommendation: a multi-channel deep fusion approach.}. In
  \bibinfo{booktitle}{\emph{IJCAI}}. \bibinfo{pages}{3805--3811}.
\newblock


\bibitem[\protect\citeauthoryear{Lin, Xie, Guan, Li, and Li}{Lin
  et~al\mbox{.}}{2014}]%
        {lin2014personalized}
\bibfield{author}{\bibinfo{person}{Chen Lin}, \bibinfo{person}{Runquan Xie},
  \bibinfo{person}{Xinjun Guan}, \bibinfo{person}{Lei Li}, {and}
  \bibinfo{person}{Tao Li}.} \bibinfo{year}{2014}\natexlab{}.
\newblock \showarticletitle{Personalized news recommendation via implicit
  social experts}.
\newblock \bibinfo{journal}{\emph{Information Sciences}}
  (\bibinfo{year}{2014}), \bibinfo{pages}{1--18}.
\newblock


\bibitem[\protect\citeauthoryear{Liu, Lian, Wang, Qiao, Chen, Sun, and Xie}{Liu
  et~al\mbox{.}}{2020}]%
        {liu2020kred}
\bibfield{author}{\bibinfo{person}{Danyang Liu}, \bibinfo{person}{Jianxun
  Lian}, \bibinfo{person}{Shiyin Wang}, \bibinfo{person}{Ying Qiao},
  \bibinfo{person}{Jiun-Hung Chen}, \bibinfo{person}{Guangzhong Sun}, {and}
  \bibinfo{person}{Xing Xie}.} \bibinfo{year}{2020}\natexlab{}.
\newblock \showarticletitle{KRED: Knowledge-aware document representation for
  news recommendations}. In \bibinfo{booktitle}{\emph{RecSys.}}
  \bibinfo{pages}{200--209}.
\newblock


\bibitem[\protect\citeauthoryear{Liu, Dolan, and Pedersen}{Liu
  et~al\mbox{.}}{2010}]%
        {liu2010personalized}
\bibfield{author}{\bibinfo{person}{Jiahui Liu}, \bibinfo{person}{Peter Dolan},
  {and} \bibinfo{person}{Elin~R{\o}nby Pedersen}.}
  \bibinfo{year}{2010}\natexlab{}.
\newblock \showarticletitle{Personalized news recommendation based on click
  behavior}. In \bibinfo{booktitle}{\emph{IUI}}. \bibinfo{pages}{31--40}.
\newblock


\bibitem[\protect\citeauthoryear{Liu, Xing, Wu, An, and Xie}{Liu
  et~al\mbox{.}}{2019}]%
        {liu2019hi}
\bibfield{author}{\bibinfo{person}{Zheng Liu}, \bibinfo{person}{Yu Xing},
  \bibinfo{person}{Fangzhao Wu}, \bibinfo{person}{Mingxiao An}, {and}
  \bibinfo{person}{Xing Xie}.} \bibinfo{year}{2019}\natexlab{}.
\newblock \showarticletitle{Hi-Fi ark: deep user representation via
  high-fidelity archive network}. In \bibinfo{booktitle}{\emph{IJCAI}}.
  \bibinfo{pages}{3059--3065}.
\newblock


\bibitem[\protect\citeauthoryear{Okura, Tagami, Ono, and Tajima}{Okura
  et~al\mbox{.}}{2017}]%
        {okura2017embedding}
\bibfield{author}{\bibinfo{person}{Shumpei Okura}, \bibinfo{person}{Yukihiro
  Tagami}, \bibinfo{person}{Shingo Ono}, {and} \bibinfo{person}{Akira Tajima}.}
  \bibinfo{year}{2017}\natexlab{}.
\newblock \showarticletitle{Embedding-based news recommendation for millions of
  users}. In \bibinfo{booktitle}{\emph{KDD}}. \bibinfo{pages}{1933--1942}.
\newblock


\bibitem[\protect\citeauthoryear{Oord, Li, and Vinyals}{Oord
  et~al\mbox{.}}{2018}]%
        {oord2018representation}
\bibfield{author}{\bibinfo{person}{Aaron van~den Oord}, \bibinfo{person}{Yazhe
  Li}, {and} \bibinfo{person}{Oriol Vinyals}.} \bibinfo{year}{2018}\natexlab{}.
\newblock \showarticletitle{Representation learning with contrastive predictive
  coding}.
\newblock \bibinfo{journal}{\emph{arXiv preprint arXiv:1807.03748}}
  (\bibinfo{year}{2018}).
\newblock


\bibitem[\protect\citeauthoryear{Pennington, Socher, and Manning}{Pennington
  et~al\mbox{.}}{2014}]%
        {pennington2014glove}
\bibfield{author}{\bibinfo{person}{Jeffrey Pennington},
  \bibinfo{person}{Richard Socher}, {and} \bibinfo{person}{Christopher
  Manning}.} \bibinfo{year}{2014}\natexlab{}.
\newblock \showarticletitle{Glove: Global vectors for word representation}. In
  \bibinfo{booktitle}{\emph{EMNLP}}. \bibinfo{pages}{1532--1543}.
\newblock


\bibitem[\protect\citeauthoryear{Qi, Wu, Wu, Huang, and Xie}{Qi
  et~al\mbox{.}}{2020}]%
        {qi2020privacy}
\bibfield{author}{\bibinfo{person}{Tao Qi}, \bibinfo{person}{Fangzhao Wu},
  \bibinfo{person}{Chuhan Wu}, \bibinfo{person}{Yongfeng Huang}, {and}
  \bibinfo{person}{Xing Xie}.} \bibinfo{year}{2020}\natexlab{}.
\newblock \showarticletitle{Privacy-Preserving News Recommendation Model
  Learning}. In \bibinfo{booktitle}{\emph{EMNLP: Findings}}.
  \bibinfo{pages}{1423--1432}.
\newblock


\bibitem[\protect\citeauthoryear{Shu, Cui, Wang, Lee, and Liu}{Shu
  et~al\mbox{.}}{2019}]%
        {shu2019defend}
\bibfield{author}{\bibinfo{person}{Kai Shu}, \bibinfo{person}{Limeng Cui},
  \bibinfo{person}{Suhang Wang}, \bibinfo{person}{Dongwon Lee}, {and}
  \bibinfo{person}{Huan Liu}.} \bibinfo{year}{2019}\natexlab{}.
\newblock \showarticletitle{defend: Explainable fake news detection}. In
  \bibinfo{booktitle}{\emph{KDD}}. \bibinfo{pages}{395--405}.
\newblock


\bibitem[\protect\citeauthoryear{Srivastava, Hinton, Krizhevsky, Sutskever, and
  Salakhutdinov}{Srivastava et~al\mbox{.}}{2014}]%
        {srivastava2014dropout}
\bibfield{author}{\bibinfo{person}{Nitish Srivastava},
  \bibinfo{person}{Geoffrey Hinton}, \bibinfo{person}{Alex Krizhevsky},
  \bibinfo{person}{Ilya Sutskever}, {and} \bibinfo{person}{Ruslan
  Salakhutdinov}.} \bibinfo{year}{2014}\natexlab{}.
\newblock \showarticletitle{Dropout: A simple way to prevent neural networks
  from overfitting}.
\newblock \bibinfo{journal}{\emph{JMLR}} (\bibinfo{year}{2014}),
  \bibinfo{pages}{1929--1958}.
\newblock


\bibitem[\protect\citeauthoryear{Vaswani, Shazeer, Parmar, Uszkoreit, Jones,
  Gomez, Kaiser, and Polosukhin}{Vaswani et~al\mbox{.}}{2017}]%
        {vaswani2017attention}
\bibfield{author}{\bibinfo{person}{Ashish Vaswani}, \bibinfo{person}{Noam
  Shazeer}, \bibinfo{person}{Niki Parmar}, \bibinfo{person}{Jakob Uszkoreit},
  \bibinfo{person}{Llion Jones}, \bibinfo{person}{Aidan~N Gomez},
  \bibinfo{person}{{\L}ukasz Kaiser}, {and} \bibinfo{person}{Illia
  Polosukhin}.} \bibinfo{year}{2017}\natexlab{}.
\newblock \showarticletitle{Attention is all you need}. In
  \bibinfo{booktitle}{\emph{NIPS}}. \bibinfo{pages}{6000--6010}.
\newblock


\bibitem[\protect\citeauthoryear{Veli{\v{c}}kovi{\'c}, Cucurull, Casanova,
  Romero, Li{\`o}, and Bengio}{Veli{\v{c}}kovi{\'c} et~al\mbox{.}}{2018}]%
        {velivckovic2017graph}
\bibfield{author}{\bibinfo{person}{Petar Veli{\v{c}}kovi{\'c}},
  \bibinfo{person}{Guillem Cucurull}, \bibinfo{person}{Arantxa Casanova},
  \bibinfo{person}{Adriana Romero}, \bibinfo{person}{Pietro Li{\`o}}, {and}
  \bibinfo{person}{Yoshua Bengio}.} \bibinfo{year}{2018}\natexlab{}.
\newblock \showarticletitle{Graph Attention Networks}. In
  \bibinfo{booktitle}{\emph{ICLR}}.
\newblock


\bibitem[\protect\citeauthoryear{Wan and Xiao}{Wan and Xiao}{2008}]%
        {wan2008single}
\bibfield{author}{\bibinfo{person}{Xiaojun Wan} {and} \bibinfo{person}{Jianguo
  Xiao}.} \bibinfo{year}{2008}\natexlab{}.
\newblock \showarticletitle{Single document keyphrase extraction using
  neighborhood knowledge.}. In \bibinfo{booktitle}{\emph{AAAI}}.
  \bibinfo{pages}{855--860}.
\newblock


\bibitem[\protect\citeauthoryear{Wang and Blei}{Wang and Blei}{2011}]%
        {wang2011collaborative}
\bibfield{author}{\bibinfo{person}{Chong Wang} {and} \bibinfo{person}{David~M
  Blei}.} \bibinfo{year}{2011}\natexlab{}.
\newblock \showarticletitle{Collaborative topic modeling for recommending
  scientific articles}. In \bibinfo{booktitle}{\emph{KDD}}.
  \bibinfo{pages}{448--456}.
\newblock


\bibitem[\protect\citeauthoryear{Wang, Wu, Liu, and Xie}{Wang
  et~al\mbox{.}}{2020}]%
        {wang2020fine}
\bibfield{author}{\bibinfo{person}{Heyuan Wang}, \bibinfo{person}{Fangzhao Wu},
  \bibinfo{person}{Zheng Liu}, {and} \bibinfo{person}{Xing Xie}.}
  \bibinfo{year}{2020}\natexlab{}.
\newblock \showarticletitle{Fine-grained interest matching for neural news
  recommendation}. In \bibinfo{booktitle}{\emph{ACL}}.
  \bibinfo{pages}{836--845}.
\newblock


\bibitem[\protect\citeauthoryear{Wang, Zhang, Xie, and Guo}{Wang
  et~al\mbox{.}}{2018}]%
        {wang2018dkn}
\bibfield{author}{\bibinfo{person}{Hongwei Wang}, \bibinfo{person}{Fuzheng
  Zhang}, \bibinfo{person}{Xing Xie}, {and} \bibinfo{person}{Minyi Guo}.}
  \bibinfo{year}{2018}\natexlab{}.
\newblock \showarticletitle{DKN: Deep knowledge-aware network for news
  recommendation}. In \bibinfo{booktitle}{\emph{WWW}}.
  \bibinfo{pages}{1835--1844}.
\newblock


\bibitem[\protect\citeauthoryear{Wu, Wu, An, Huang, Huang, and Xie}{Wu
  et~al\mbox{.}}{2019b}]%
        {wu2019ijcai}
\bibfield{author}{\bibinfo{person}{Chuhan Wu}, \bibinfo{person}{Fangzhao Wu},
  \bibinfo{person}{Mingxiao An}, \bibinfo{person}{Jianqiang Huang},
  \bibinfo{person}{Yongfeng Huang}, {and} \bibinfo{person}{Xing Xie}.}
  \bibinfo{year}{2019}\natexlab{b}.
\newblock \showarticletitle{Neural news recommendation with attentive
  multi-view learning}. In \bibinfo{booktitle}{\emph{IJCAI}}.
  \bibinfo{pages}{3863--3869}.
\newblock


\bibitem[\protect\citeauthoryear{Wu, Wu, An, Huang, Huang, and Xie}{Wu
  et~al\mbox{.}}{2019c}]%
        {wu2019npa}
\bibfield{author}{\bibinfo{person}{Chuhan Wu}, \bibinfo{person}{Fangzhao Wu},
  \bibinfo{person}{Mingxiao An}, \bibinfo{person}{Jianqiang Huang},
  \bibinfo{person}{Yongfeng Huang}, {and} \bibinfo{person}{Xing Xie}.}
  \bibinfo{year}{2019}\natexlab{c}.
\newblock \showarticletitle{Npa: Neural news recommendation with personalized
  attention}. In \bibinfo{booktitle}{\emph{KDD}}. \bibinfo{pages}{2576--2584}.
\newblock


\bibitem[\protect\citeauthoryear{Wu, Wu, An, Huang, and Xie}{Wu
  et~al\mbox{.}}{2019a}]%
        {wutanr}
\bibfield{author}{\bibinfo{person}{Chuhan Wu}, \bibinfo{person}{Fangzhao Wu},
  \bibinfo{person}{Mingxiao An}, \bibinfo{person}{Yongfeng Huang}, {and}
  \bibinfo{person}{Xing Xie}.} \bibinfo{year}{2019}\natexlab{a}.
\newblock \showarticletitle{Neural News Recommendation with Topic-Aware News
  Representation}. In \bibinfo{booktitle}{\emph{ACL}}.
  \bibinfo{pages}{1154--1159}.
\newblock


\bibitem[\protect\citeauthoryear{Wu, Wu, An, Qi, Huang, Huang, and Xie}{Wu
  et~al\mbox{.}}{2019d}]%
        {wu2019neurald}
\bibfield{author}{\bibinfo{person}{Chuhan Wu}, \bibinfo{person}{Fangzhao Wu},
  \bibinfo{person}{Mingxiao An}, \bibinfo{person}{Tao Qi},
  \bibinfo{person}{Jianqiang Huang}, \bibinfo{person}{Yongfeng Huang}, {and}
  \bibinfo{person}{Xing Xie}.} \bibinfo{year}{2019}\natexlab{d}.
\newblock \showarticletitle{Neural news recommendation with heterogeneous user
  behavior}. In \bibinfo{booktitle}{\emph{EMNLP}}. \bibinfo{pages}{4876--4885}.
\newblock


\bibitem[\protect\citeauthoryear{Wu, Wu, Ge, Qi, Huang, and Xie}{Wu
  et~al\mbox{.}}{2019e}]%
        {wu2019neuralc}
\bibfield{author}{\bibinfo{person}{Chuhan Wu}, \bibinfo{person}{Fangzhao Wu},
  \bibinfo{person}{Suyu Ge}, \bibinfo{person}{Tao Qi},
  \bibinfo{person}{Yongfeng Huang}, {and} \bibinfo{person}{Xing Xie}.}
  \bibinfo{year}{2019}\natexlab{e}.
\newblock \showarticletitle{Neural news recommendation with multi-head
  self-attention}. In \bibinfo{booktitle}{\emph{EMNLP}}.
  \bibinfo{pages}{6390--6395}.
\newblock


\bibitem[\protect\citeauthoryear{Wu, Wu, Huang, and Xie}{Wu
  et~al\mbox{.}}{2020b}]%
        {wu2020ccf}
\bibfield{author}{\bibinfo{person}{Chuhan Wu}, \bibinfo{person}{Fangzhao Wu},
  \bibinfo{person}{Yongfeng Huang}, {and} \bibinfo{person}{Xing Xie}.}
  \bibinfo{year}{2020}\natexlab{b}.
\newblock \showarticletitle{Neural news recommendation with negative feedback}.
\newblock \bibinfo{journal}{\emph{CCF TPCI}} (\bibinfo{year}{2020}),
  \bibinfo{pages}{178--188}.
\newblock


\bibitem[\protect\citeauthoryear{Wu, Wu, Liu, and Huang}{Wu
  et~al\mbox{.}}{2019f}]%
        {wu2019hierarchical}
\bibfield{author}{\bibinfo{person}{Chuhan Wu}, \bibinfo{person}{Fangzhao Wu},
  \bibinfo{person}{Junxin Liu}, {and} \bibinfo{person}{Yongfeng Huang}.}
  \bibinfo{year}{2019}\natexlab{f}.
\newblock \showarticletitle{Hierarchical user and item representation with
  three-tier attention for recommendation}. In
  \bibinfo{booktitle}{\emph{NAACL}}. \bibinfo{pages}{1818--1826}.
\newblock


\bibitem[\protect\citeauthoryear{Wu, Wu, Qi, and Huang}{Wu
  et~al\mbox{.}}{2020c}]%
        {wu2020clickbait}
\bibfield{author}{\bibinfo{person}{Chuhan Wu}, \bibinfo{person}{Fangzhao Wu},
  \bibinfo{person}{Tao Qi}, {and} \bibinfo{person}{Yongfeng Huang}.}
  \bibinfo{year}{2020}\natexlab{c}.
\newblock \showarticletitle{Clickbait Detection with Style-Aware Title Modeling
  and Co-attention}. In \bibinfo{booktitle}{\emph{CCL}}.
  \bibinfo{pages}{430--443}.
\newblock


\bibitem[\protect\citeauthoryear{Wu, Wu, Qi, and Huang}{Wu
  et~al\mbox{.}}{2020d}]%
        {wu2020sentirec}
\bibfield{author}{\bibinfo{person}{Chuhan Wu}, \bibinfo{person}{Fangzhao Wu},
  \bibinfo{person}{Tao Qi}, {and} \bibinfo{person}{Yongfeng Huang}.}
  \bibinfo{year}{2020}\natexlab{d}.
\newblock \showarticletitle{SentiRec: Sentiment Diversity-aware Neural News
  Recommendation}. In \bibinfo{booktitle}{\emph{AACL}}.
  \bibinfo{pages}{44--53}.
\newblock


\bibitem[\protect\citeauthoryear{Wu, Wu, Qi, and Huang}{Wu
  et~al\mbox{.}}{2020e}]%
        {wuuser}
\bibfield{author}{\bibinfo{person}{Chuhan Wu}, \bibinfo{person}{Fangzhao Wu},
  \bibinfo{person}{Tao Qi}, {and} \bibinfo{person}{Yongfeng Huang}.}
  \bibinfo{year}{2020}\natexlab{e}.
\newblock \showarticletitle{User modeling with click preference and reading
  satisfaction for news recommendation}. In \bibinfo{booktitle}{\emph{IJCAI}}.
  \bibinfo{pages}{3023--3029}.
\newblock


\bibitem[\protect\citeauthoryear{Wu, Wu, Qi, Lian, Huang, and Xie}{Wu
  et~al\mbox{.}}{2020f}]%
        {wu2020ptum}
\bibfield{author}{\bibinfo{person}{Chuhan Wu}, \bibinfo{person}{Fangzhao Wu},
  \bibinfo{person}{Tao Qi}, \bibinfo{person}{Jianxun Lian},
  \bibinfo{person}{Yongfeng Huang}, {and} \bibinfo{person}{Xing Xie}.}
  \bibinfo{year}{2020}\natexlab{f}.
\newblock \showarticletitle{PTUM: Pre-training User Model from Unlabeled User
  Behaviors via Self-supervision}. In \bibinfo{booktitle}{\emph{EMNLP:
  Findings}}. \bibinfo{pages}{1939--1944}.
\newblock


\bibitem[\protect\citeauthoryear{Wu, Wu, Wang, Huang, and Xie}{Wu
  et~al\mbox{.}}{2021}]%
        {wu2020fairness}
\bibfield{author}{\bibinfo{person}{Chuhan Wu}, \bibinfo{person}{Fangzhao Wu},
  \bibinfo{person}{Xiting Wang}, \bibinfo{person}{Yongfeng Huang}, {and}
  \bibinfo{person}{Xing Xie}.} \bibinfo{year}{2021}\natexlab{}.
\newblock \showarticletitle{FairRec:Fairness-aware News Recommendation with
  Decomposed Adversarial Learning}. In \bibinfo{booktitle}{\emph{AAAI}}.
\newblock


\bibitem[\protect\citeauthoryear{Wu, Qiao, Chen, Wu, Qi, Lian, Liu, Xie, Gao,
  Wu, et~al\mbox{.}}{Wu et~al\mbox{.}}{2020a}]%
        {wu2020mind}
\bibfield{author}{\bibinfo{person}{Fangzhao Wu}, \bibinfo{person}{Ying Qiao},
  \bibinfo{person}{Jiun-Hung Chen}, \bibinfo{person}{Chuhan Wu},
  \bibinfo{person}{Tao Qi}, \bibinfo{person}{Jianxun Lian},
  \bibinfo{person}{Danyang Liu}, \bibinfo{person}{Xing Xie},
  \bibinfo{person}{Jianfeng Gao}, \bibinfo{person}{Winnie Wu}, {et~al\mbox{.}}}
  \bibinfo{year}{2020}\natexlab{a}.
\newblock \showarticletitle{MIND: A large-scale dataset for news
  recommendation}. In \bibinfo{booktitle}{\emph{ACL}}.
  \bibinfo{pages}{3597--3606}.
\newblock


\bibitem[\protect\citeauthoryear{Zheng, Zhang, Zheng, Xiang, Yuan, Xie, and
  Li}{Zheng et~al\mbox{.}}{2018}]%
        {zheng2018drn}
\bibfield{author}{\bibinfo{person}{Guanjie Zheng}, \bibinfo{person}{Fuzheng
  Zhang}, \bibinfo{person}{Zihan Zheng}, \bibinfo{person}{Yang Xiang},
  \bibinfo{person}{Nicholas~Jing Yuan}, \bibinfo{person}{Xing Xie}, {and}
  \bibinfo{person}{Zhenhui Li}.} \bibinfo{year}{2018}\natexlab{}.
\newblock \showarticletitle{DRN: A deep reinforcement learning framework for
  news recommendation}. In \bibinfo{booktitle}{\emph{WWW}}.
  \bibinfo{pages}{167--176}.
\newblock


\bibitem[\protect\citeauthoryear{Zhu, Zhou, Song, Tan, and Li}{Zhu
  et~al\mbox{.}}{2019}]%
        {danzhu2019}
\bibfield{author}{\bibinfo{person}{Qiannan Zhu}, \bibinfo{person}{Xiaofei
  Zhou}, \bibinfo{person}{Zeliang Song}, \bibinfo{person}{Jianlong Tan}, {and}
  \bibinfo{person}{Guo Li}.} \bibinfo{year}{2019}\natexlab{}.
\newblock \showarticletitle{DAN: Deep attention neural network for news
  recommendation}. In \bibinfo{booktitle}{\emph{AAAI}}.
  \bibinfo{pages}{5973--5980}.
\newblock


\end{thebibliography}


\end{document}